\documentclass[%
 reprint,
 amsmath,amssymb,
 aps,
]{revtex4-2}

\usepackage{graphicx}
\usepackage{dcolumn}
\usepackage{bm}
\usepackage{braket}
\usepackage{xcolor}
\setcitestyle{numbers,square,comma,sort&compress}

\usepackage{hyperref}
\hypersetup{colorlinks=true, citecolor=blue, linkcolor=blue, urlcolor=blue}

\begin{document}

\preprint{APS/123-QED}

%
\title{Robust Nonclassical magnon pair generation and Cauchy-Schwarz inequality violation in a hybrid electromagnonic system}

\author{Sabur A Barbhuiya}
\email{sabur07@rnd.iitg.ac.in}
 \author{Ajeet Kumar}
 \email{ajeetk1005@iitg.ac.in}
\author{Amarendra K Sarma}%
 \email{aksarma@iitg.ac.in}
\affiliation{%
 Department of Physics, Indian Institute of Technology Guwahati, Guwahati- 781039, Assam,  India\\
}%

\begin{abstract}
Magnonic systems are a promising resource for quantum technologies because of
their ability to couple significantly disparate quantum platforms and the
generation of nonclassical magnon states through magnon blockade has attracted
growing attention. In practice, however magnon blockade relies on a
destructive interference that is easily spoiled by fabrication-induced coupling
asymmetries. Here we show that the relative phase between two magnon drives acts
as an active compensation knob that restores this interference even when the
couplings are mismatched. We study two Kittel magnon modes in a hybrid system
in which a superconducting qubit coupled to a common cavity mode mediates the
inter-mode interaction and we find that tuning the drive phase produces both
magnon blockade and a strong violation of the classical Cauchy--Schwarz
inequality. We derive the analytic condition for the phase that compensates a
given coupling asymmetry and confirm it against exact numerical simulations. The
drive phase thereby serves as a control knob that switches the system between
classical and quantum regimes. Our results enable phase-controlled magnonic quantum information processing.
\end{abstract}

\maketitle
%
%
%
%
%
%
%

\section{Introduction}

In recent years, hybrid quantum systems\cite{lachance2019hybrid} have attracted significant attention due to their potential to combine the unique advantages of different physical platforms for quantum information processing\cite{QuantumInformationProcessing_1,QuantumInformationProcessing_2,QuantumInformationProcessing_3}, quantum spatial networking\cite{QuantumNetworking_1,QuantumNetworking_2}, dark modes, tripartile entanglement\cite{TripartileEntanglement_1,TripartileEntanglement_2,Magnon-photon-phonon_entanglement},enhanced sensing\cite{QuantumSensing_1,SensingEntanglement,QuantumSensing_2}, precision metrology\cite{Metrology_1,Metrology_2} and blockade\cite{MagnonBlockade_1,MagnonBlockade_2,EntanglementMagnonBlockade_3} effects. Among these diverse platforms magnonics is based on the collective spin excitations (magnons) in ferromagnetic insulators like yttrium-iron-garnet (YIG)\cite{YIG_1,YIG_2SpinDensity_1} spheres has emerged as a highly promising candidate because of their exceptionally high spin density\cite{YIG_2SpinDensity_1,SpinDensity_2}, low intrinsic dissipation\cite{Dissipations_1,Dissipations_2,QuantumInformationProcessing_2}, and excellent geometric tunability. Macroscopic YIG spheres exhibit strong coupling capabilities with various quantum systems including microwave cavity photons\cite{MicrowaveCavity_1,MicrowaveCavity_2,MicrowaveCavity_3}, optical photons\cite{Magnon-photon-phonon_entanglement}, phonons\cite{Magnon-photon-phonon_entanglement,phonon_2}, and superconducting qubits\cite{SuperconductingQubit_1,SuperconductingQubit_2}. In particular to integrate macroscopic YIG spheres with highly nonlinear superconducting transmon qubits within a three dimensional cavity quantum electrodynamics (cQED) architecture has opened new avenues for exploring macroscopic quantum phenomena\cite{SuperconductingQubit_1,SuperconductingQubit_2}. In these setups the microwave cavity acts as a robust and coherent quantum data bus that it is mediating long range interactions between spatially separated components without requiring direct physical contact\cite{MicrowaveCavity_3,MagnonBlockade_2}.
 
A fundamental milestone in the quantum control of such hybrid systems is the realization of the magnon blockade which is an effect strictly analogous to the well-known photon blockade in quantum optics\cite{PhotonBlockade_1,PhotonBlockade_2,PhotonBlockade_3,PhotonBlockade_4}. In a magnon blockade the excitation of a single magnon within the system dynamically shifts the energy levels which physically prevent the subsequent generation of a second magnon. This serves as a critical mechanism for the deterministic generation of single magnon states which are essential resources for quantum communication and quantum simulation. The conventional magnon blockade (CMB)\cite{MagnonBlockade_1,ConventionalUnconventional} fundamentally relies on strong inherent anharmonicity in the energy spectrum which is often induced by coupling to a nonlinear qubit that requiring the system to operate deep within the strong coupling regime. However for achieving and maintaining this strong coupling is experimentally demanding and highly susceptible to environmental decoherence. To overcome this strict limitation, the unconventional magnon blockade (UMB) \cite{MagnonBlockade_2,ConventionalUnconventional,Unconventional_2,Unconventional_3} has been proposed and actively investigated. The UMB achieves strong sub-Poissonian statistics even in the weak or moderate coupling regimes by utilizing destructive quantum interference between distinct multi level excitation pathways. By carefully designing these pathways the transition probability to the two excitation state can be completely canceled \cite{MagnonBlockade_2}.

While early theoretical and experimental efforts have successfully demonstrated magnon blockades in hybrid systems containing a single YIG sphere coupled to a superconducting qubit \cite{MagnonBlockade_1,SuperconductingQubit_2} and the exploration of mult mode magnon architectures is rapidly expanding\cite{RapidlyExpanding_1}. Multimode systems\cite{multipartite_1,MagnonBlockade_2} offer richer interference dynamics and the potential for generating correlated states\cite{generatingCorrelatedPairs_1,Magnon-photon-phonon_entanglement} across macroscopic distances. For instance, recent studies have proposed using two YIG spheres and a single transmon qubit to generate quantum correlated, nonclassical magnon pairs and observe macroscopic violations of the Cauchy-Schwarz inequality (CSI)\cite{MagnonBlockade_2,CSI_Violation_1,CsiViolation}. These studies have shown that by utilizing virtual photon excitation mediated by a cavity mode the effective couplings can be established not only between the Kittel modes and the qubit but also indirectly between the two Kittel modes themselves. Furthermore by optimizing relative phase and driving strength ratios between the system components that has been shown significant enhancement to the blockade effect. However the use of externally applied highly tunable relative driving phases as the primary active control knob which is specifically relying on the interference between directly driven macroscopic magnon modes while leaving the auxiliary qubit undriven remains a largely unexplored frontier. Simplifying the active drive architecture while maximizing the blockade fidelity and pair correlations is highly desirable for scalable experimental implementations.

In this paper, we theoretically propose and comprehensively investigate an actively driven tripartite hybrid quantum system consisting of two YIG spheres and a transmon qubit mediated by a common microwave cavity bus. By strictly operating in the large detuning dispersive regime we can adiabatically eliminate the cavity mode to establish effective direct dipole dipole couplings between the qubit and the Kittel modes as well as an indirect cross coupling between the two magnons. Crucially, we apply external microwave drives exclusively to the two macroscopic magnon modes, introducing a highly tunable relative phase ($\phi$) between them while leaving the transmon qubit completely undriven. We derive the effective Hamiltonian analytically and solve the Lindblad master equation for the open quantum system numerically. We demonstrate that this relative phase acts as a powerful, dynamic control knob to regulate the destructive interference pathways. We evaluate the equal time second order autocorrelation and cross correlation functions to show that our simplified driving architecture can simultaneously achieve a strong unconventional single magnon blockade. Furthermore, we definitively prove the generation of highly correlated nonclassical magnon pairs that unambiguously violate the macroscopic Cauchy-Schwarz inequality. Our findings provide a feasible and robust pathway for single magnon generation and continuous variable quantum correlation in multimode magnonic networks.
\section{Model and Hamiltonian of The Hybrid Quantum Architecture}

We consider a tripartite hybrid quantum system consisting of two single crystalline yttrium-iron-garnet (YIG) spheres and a transmon type superconducting qubit where everything is combined within a three-dimensional microwave cavity. The uniform spin precessions within the YIG spheres which are known as Kittel modes are treated as quantum harmonic oscillators and are represented by the bosonic annihilation (creation) operators $m_1$ ($m_1^\dagger$) and $m_2$ ($m_2^\dagger$). The superconducting transmon qubit is approximated as an effective two-level system that is described by the Pauli operators $\sigma_z$, $\sigma_+$, and $\sigma_-$.

The two Kittel modes and the transmon qubit are placed near the antinodes of the cavity's magnetic and electric fields respectively allowing them to couple to the single cavity mode $c$ ($c^\dagger$) via magnetic and electric dipole interactions, as shown in Fig.1. We assume that the direct interactions between the qubit and the YIG spheres as well as between the two YIG spheres themselves are negligible \cite{MagnonBlockade_2,DirectCouplingBetweenMagnonandQubit_1,DirectCouplingBetweenMagnonandQubit_2} due to their spatial separation. Instead the effective couplings among the qubit and the two Kittel modes are established entirely via virtual photon excitation mediated by the cavity mode. Crucially, such manipulation involving virtual quanta is physically feasible and actively allowed by current experiments\cite{SuperconductingQubit_2,SuperconductingQubit_1}.

\begin{figure*}[htbp]
\centering
\includegraphics[width=0.60\textwidth]{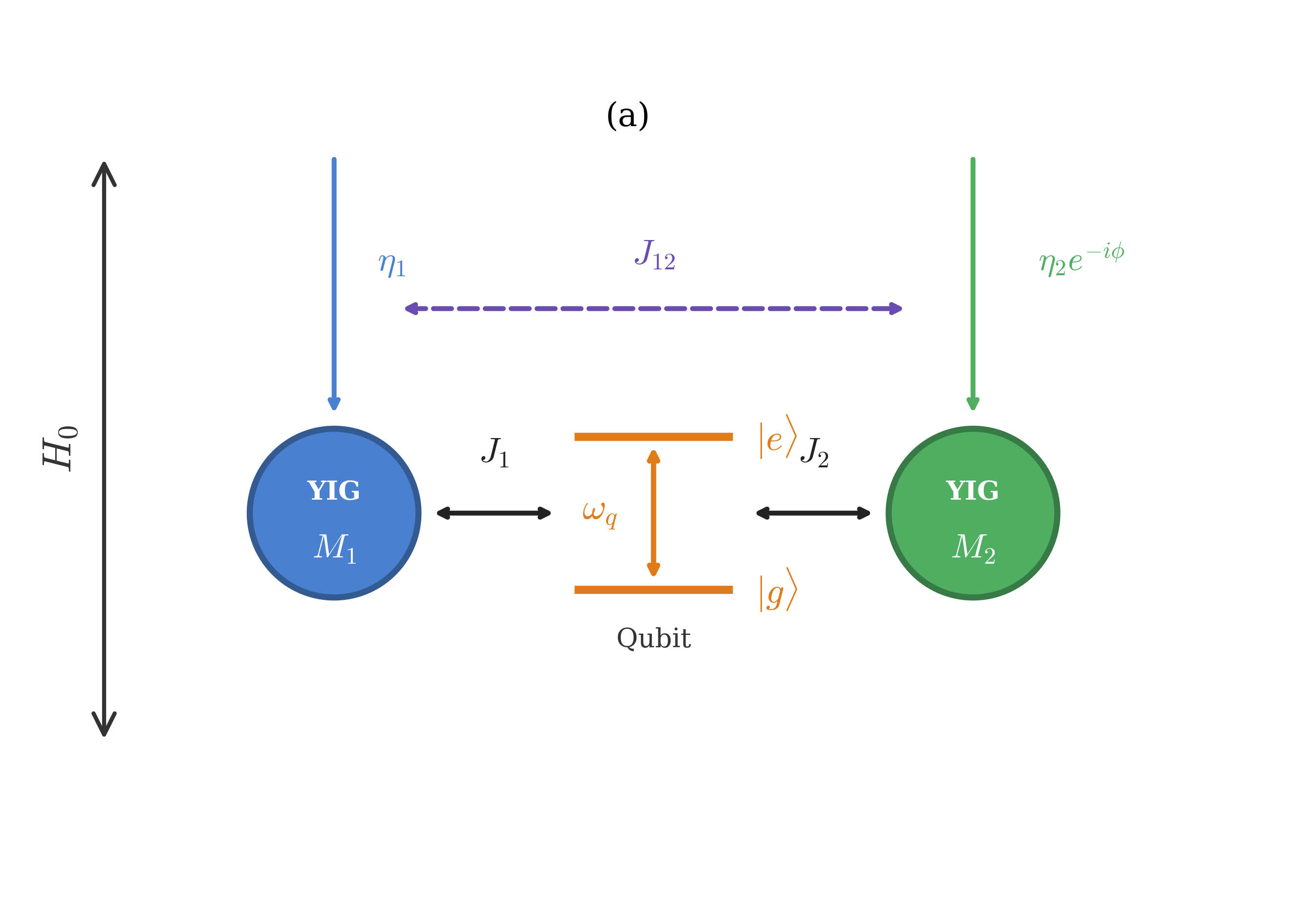}\\ \includegraphics[width=0.48\textwidth]{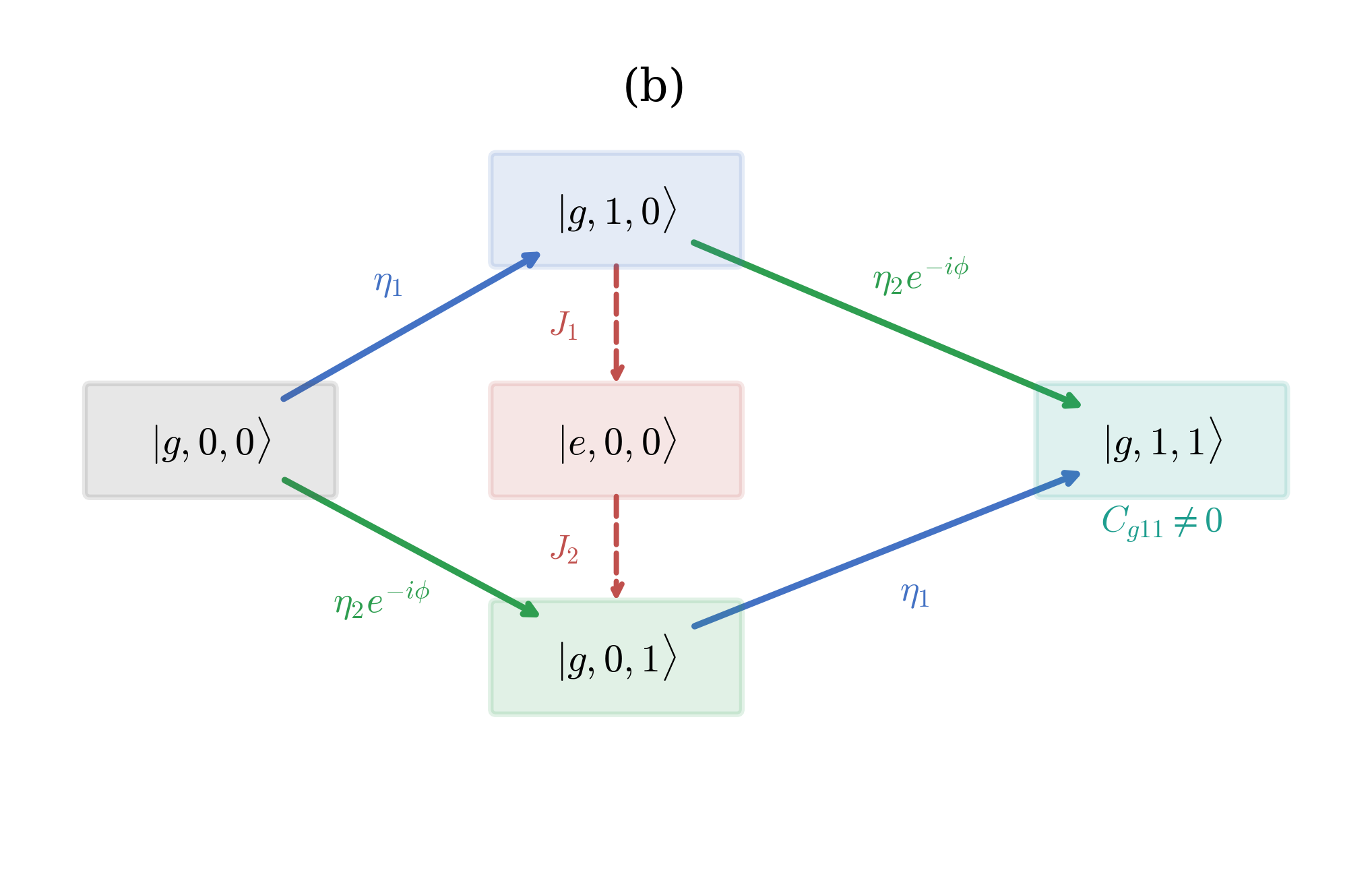} \includegraphics[width=0.48\textwidth]{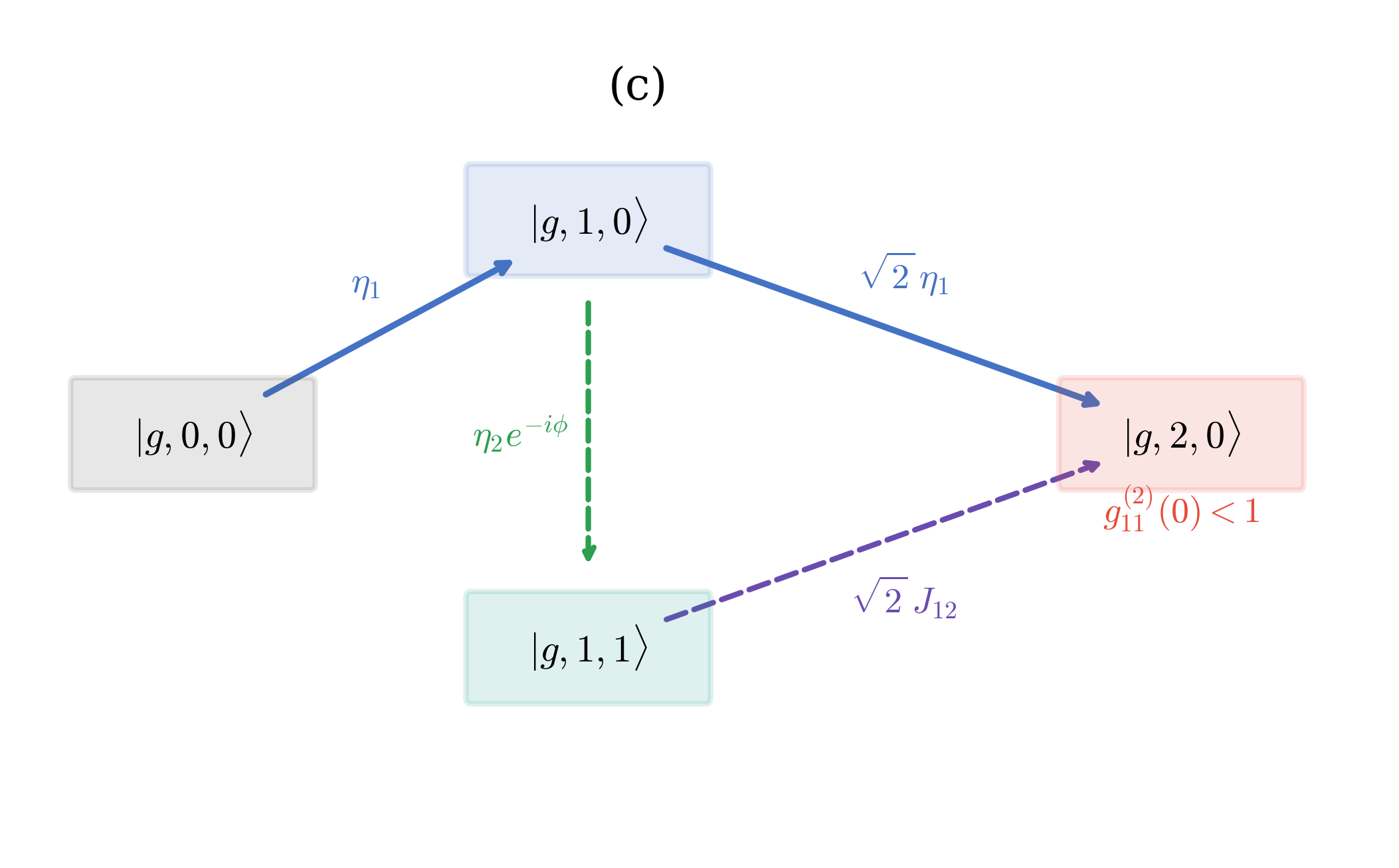}\\
\caption{Hybrid magnon--qubit system and the interference mechanism
underlying the two-magnon blockade.
(a)~Effective-model configuration [Eq.~(3)] which is obtained after adiabatic
elimination of the microwave cavity.  Two Kittel magnon modes $M_1$ and $M_2$
and a superconducting qubit (whose transition frequency is$\omega_q$) are directly
coupled by the cavity-induced interactions such as magnon--qubit couplings $J_1$ and
$J_2$ and an effective magnon--magnon coupling $J_{12}$.  A static field
$H_0$ sets the magnon frequencies and the two modes are driven by $\eta_1$
and $\eta_2 e^{-i\phi}$ with $\phi$ is the relative drive phase.
(b)~Excitation pathways to the cross-mode two-magnon state
$|g,1,1\rangle$.  The two direct-drive paths (solid) each carrying one factor
$\eta_1$ and one factor $\eta_2 e^{-i\phi}$ together with the qubit-mediated
path through $|e,0,0\rangle$ (dashed, $J_1J_2$) add as constructively, so that
$|g,1,1\rangle$ remains occupied that is $C_{g11}\neq0$.
(c)~Excitation pathways to the same-mode two-magnon state
$|g,2,0\rangle$.  The direct double-drive path (solid, $\eta_1\to\sqrt{2}\,\eta_1$)
interferes with the cavity-induced path through $|g,1,1\rangle$
(dashed, $\eta_2 e^{-i\phi}$ then $\sqrt{2}\,J_{12}$). Also, near the optimal phase
$\phi^{*}\simeq1.18\pi$ this interference is destructive by suppressing $C_{g20}$
so that the mode is antibunched that is $g^{(2)}_{11}(0)<1$ and the same mechanism
suppresses $C_{g02}$.  The combination of suppressed same-mode amplitudes and
an enhanced cross-mode amplitude is what generates the strong nonclassical
cross-correlation and the large Cauchy-Schwarz violation $R\gg1$ under strong
driving.}
\label{fig:1}
\end{figure*}

To dynamically manipulate the quantum interference within the system external microwave fields are applied to actively drive both the magnon modes. We introduce local highly tunable phase to driving field. As recent studies have demonstrated, these relative phase serves as active control knob to regulate destructive interference pathways\cite{wu2021phase,PhysRevA.110.012459,ConventionalUnconventional}, paving the way for unconventional blockade effects without requiring strong inherent anharmonicity.
Setting $\hbar = 1$, the total Hamiltonian of the system in the laboratory frame is given by
\begin{equation}
    H_{lab} = H_{free} + H_{int} + H_{drive}
    \tag{1}
\end{equation}
The free energy Hamiltonian of the uncoupled modes is given by
\begin{equation}
    H_{free} = \omega_c c^\dagger c + \sum_{j=1,2} \omega_{mj} m_j^\dagger m_j + \omega_q \sigma_+ \sigma_-
    \tag{2a}
\end{equation}
where $\omega_c$, $\omega_{mj}$, and $\omega_q$ represent the resonance frequencies of the microwave cavity, the two magnon modes, and the qubit respectively. The interaction Hamiltonian describing the direct coupling between the local modes and the cavity bus is given by
\begin{equation}
    H_{int} = \sum_{j=1,2} g_{jc} (c^\dagger m_j + c m_j^\dagger) + g_{qc} (c^\dagger \sigma_- + c \sigma_+)
    \tag{2b}
\end{equation}
where $g_{jc}$ and $g_{qc}$ denote the direct coupling strengths of the respective Kittel modes and the qubit to the cavity field. Finally, the driving Hamiltonian is expressed as 
\begin{equation}
\begin{split}
H_{drive} =\ & \eta_1 \left( m_1^\dagger e^{-i\omega_d t} + m_1 e^{i\omega_d t} \right) \\
& + \eta_2 \left( m_2^\dagger e^{-i(\omega_d t + \phi)} + m_2 e^{i(\omega_d t + \phi)} \right)
\end{split}
\tag{2c}
\end{equation}
where $\omega_d$ is the frequency of the external microwave drives and $\eta_1$ and $\eta_2$ represent their respective driving amplitudes.  To remove the explicit time dependence introduced by the external microwave drives we apply a unitary transformation $U = \exp[-i\omega_d t(c^\dagger c + \sum_{j=1,2} m_j^\dagger m_j + \sigma_+ \sigma_-)]$ to shift the system into the rotating frame.

To achieve coherent control without populating the cavity, we operate strictly in the large detuning (dispersive) limit\cite{MagnonBlockade_2}. The detuning between the cavity mode and the external drives ($\Delta_c = \omega_c - \omega_d$) is assumed to be much larger than the respective direct coupling strengths and drive amplitudes ($|\Delta_c| \gg \{g_{1c}, g_{2c}, g_{qc}, \eta_1, \eta_2\}$)\cite{MagnonBlockade_2}. Under this condition, the cavity mode is virtually unpopulated with real photons ($\langle c^\dagger c \rangle \approx 0$)\cite{MagnonBlockade_2}.  The system dynamics is governed by the virtual exchange of cavity photons because the direct couplings are far off-resonant. To analytically isolate this effect, we applied the Fröhlich Nakajima transformation to adiabatically eliminate the cavity mode from the dynamics. This transformation replaces the cavity mediated interactions with effective direct dipole-dipole couplings between the local modes. The resulting effective time-independent Hamiltonian becomes as follows,

\begin{equation}
\begin{split}
H_{eff} ={}& \sum_{j=1,2} \tilde{\Delta}_{m_j} m_j^\dagger m_j + \tilde{\Delta}_q \sigma_+ \sigma_-  + J_{12}(m_1^\dagger m_2 + m_2^\dagger m_1) \\
& + \sum_{j=1,2} J_{j} (m_j^\dagger \sigma_- + m_j \sigma_+)  + \eta_1 (m_1^\dagger + m_1) \\
& + \eta_2 \left( m_2^\dagger e^{-i\phi} + m_2 e^{i\phi} \right)
\end{split}
\tag{3}
\end{equation}

Here, the bare detunings of the modes from the drive frequency are defined as $\Delta_{m_j} = \omega_{m_j} - \omega_d$ and $\Delta_q = \omega_q - \omega_d$, while the relative detunings between the cavity and the local modes are defined as $\delta_{cm_j} = \Delta_c - \Delta_{m_j}$ and $\delta_{cq} = \Delta_c - \Delta_q$. Using these the Stark shifted effective detunings of the magnons and the qubit are accurately given by $\tilde{\Delta}_{m_j} = \Delta_{m_j} - \frac{J_j^2}{\delta_{cm_j}}$ and $\tilde{\Delta}_q = \Delta_q - \frac{g_q^2}{\delta_{cq}}$\cite{MagnonBlockade_2}. Crucially, the adiabatic elimination yields the exact effective qubit-magnon coupling strengths $J_{j} = -\frac{1}{2} g_{jc} g_{qc} \left( \frac{1}{\delta_{cm_j}} + \frac{1}{\delta_{cq}} \right)$\cite{MagnonBlockade_2,SuperconductingQubit_2,SuperconductingQubit_1} and the indirect magnon magnon cross-coupling $J_{12} = -\frac{1}{2} g_{1c} g_{2c} \left( \frac{1}{\delta_{cm_1}} + \frac{1}{\delta_{cm_2}} \right)$.

To describe the realistic dynamics of our hybrid quantum system interacting with its environment we must account for energy dissipation. We introduce the dissipation rates for each mode as $\gamma_j$ for the magnon decay rate of the $j$-th Kittel mode, and $\Gamma_q$ for the relaxation rate of the transmon qubit. The dissipative nature of the system can be phenomenologically described by introducing a non-Hermitian effective Hamiltonian given by
\begin{equation}
\hat{H}_{non-Herm} = \hat{H}_{eff} - i \sum_{j=1,2} \frac{\gamma_j}{2} m_j^\dagger m_j - i \frac{\Gamma_q}{2} \sigma_+ \sigma_-
\tag{4}
\end{equation}
which effectively corresponds to making the Stark shifted detunings complex to form $\tilde{\Delta}_{m_j} \rightarrow \tilde{\Delta}_{m_j} - i \gamma_j / 2$ and $\tilde{\Delta}_q \rightarrow \tilde{\Delta}_q - i \Gamma_q / 2$.
Assuming a near zero temperature environment which is highly accurate for such systems operating in dilution refrigerators at around millikelvin temperatures\cite{MicrowaveCavity_2,lachance2019hybrid} the full time evolution of the system's density matrix $\rho$ is rigorously governed by the Lindblad master equation:
\begin{equation}
\frac{d\hat{\rho}}{dt} = -i[\hat{H}_{eff}, \hat{\rho}] + \sum_{j=1,2} \gamma_j \mathcal{D}[\hat{m}_j]\hat{\rho} + \Gamma_q \mathcal{D}[\sigma_-]\hat{\rho}
\tag{5}
\end{equation}

where the standard Lindblad superoperator for a generic collapse operator $\hat{O}$ is defined as $\mathcal{D}[\hat{O}]\hat{\rho} = \hat{O} \hat{\rho} \hat{O}^\dagger - \frac{1}{2} \{\hat{O}^\dagger \hat{O}, \hat{\rho}\}$.\\

By numerically solving Eq. (5) for the steady state density matrix $\rho_s$ (i.e., setting $d\hat{\rho}/dt = 0$), we can evaluate the statistical properties of the magnon modes. The degree of magnon blockade within each individual Kittel mode is characterized by the normalized, equal time second order autocorrelation function \cite{Correlations_1,MagnonBlockade_2,MagnonBlockade_1,ConventionalUnconventional}  as well as cross-correlation function\cite{MagnonBlockade_2} given by 
\begin{equation}
g_{aa}^{(2)}(0) = \frac{\text{Tr}(\rho_s m_a^\dagger m_a^\dagger m_a m_a)}{[\text{Tr}(\rho_s m_a^\dagger m_a)]^2} \quad \text{for } a \in {1, 2}
\tag{6}
\end{equation}

\begin{equation}
g_{12}^{(2)}(0) = \frac{\text{Tr}(\rho_s m_1^\dagger m_2^\dagger m_2 m_1)}{\text{Tr}(\rho_s m_1^\dagger m_1)\text{Tr}(\rho_s m_2^\dagger m_2)}
\tag{7}
\end{equation}

The equal time second order autocorrelation function \( g^{(2)}_{aa}(0) \) serves as a key indicator of magnon statistics. A value of  \( g^{(2)}_{aa}(0) > 1 \) signifies that the magnons follow a super-Poissonian distribution which exhibits classical bunching behavior. If \( g^{(2)}_{aa}(0) = 1 \), the statistics become Poissonian which corresponds to a coherent state that lies at the boundary between the classical and quantum regimes. When \( g^{(2)}_{aa}(0) < 1 \), the magnons obey sub-Poissonian statistics  (nonclassical antibunching). As \( g^{(2)}_{aa}(0) \) approaches zero, complete magnon blockade occurs which means that magnons repel each other and it can be exploited to create a single magnon source .

A value substantially smaller than unity for the equal time second order cross correlation function $g^{(2)}_{12}(0)$ suggests that the state $|g, 1, 1 \rangle$ seems unlikely to be populated. As a result, there is anti correlation between the two modes since excitation of the first Kittel mode prevents stimulation of the second \cite{PhysRevA.94.043814}. On the other hand, the emission of two mode magnon pairs is indicated by ($ g^{(2)}_{12}(0) > 1 $). It is crucial to remember that it does not directly verify quantum correlations between the two Kittel modes, even in cases where $g^{(2)}_{12}(0) < 1$. To develop a precise quantum behavior, additional constraints must be applied especially the Cauchy-Schwarz inequality (CSI). So, in the classical regime the cross correlation is strictly bounded by the geometric mean of the autocorrelations. We quantify this boundary by defining the CSI violation factor $R$ which is given by \cite{PhysRevA.105.043711, PhysRevA.90.052111}
\begin{equation}
R = \frac{g_{12}^{(2)}(0)}{\sqrt{g_{11}^{(2)}(0) g_{22}^{(2)}(0)}}
\tag{8}
\end{equation}
When $R \le 1$ is the correlations between the two magnon modes can be fully described by classical physics. However, a value of $R > 1$ firmly violates the classical CSI boundary which provides unambiguous macroscopic evidence of strongly correlated nonclassical magnon pairs generated via our actively driven quantum interference mechanism.

\section{Results and Discussions}

In this section, we look at numerous features of equal-time second-order autocorrelation $g_{11}$ and $g_{22}$, cross correlation $g_{12}$, and the CSI violation factor (R) based on various parameters such as the driving and coupling ratios of the first and second Kittle modes, as well as the qubit. The results are shown in detail in the following figures. 

%

\begin{figure}[htbp]
\centering
\includegraphics[width=0.48\textwidth]{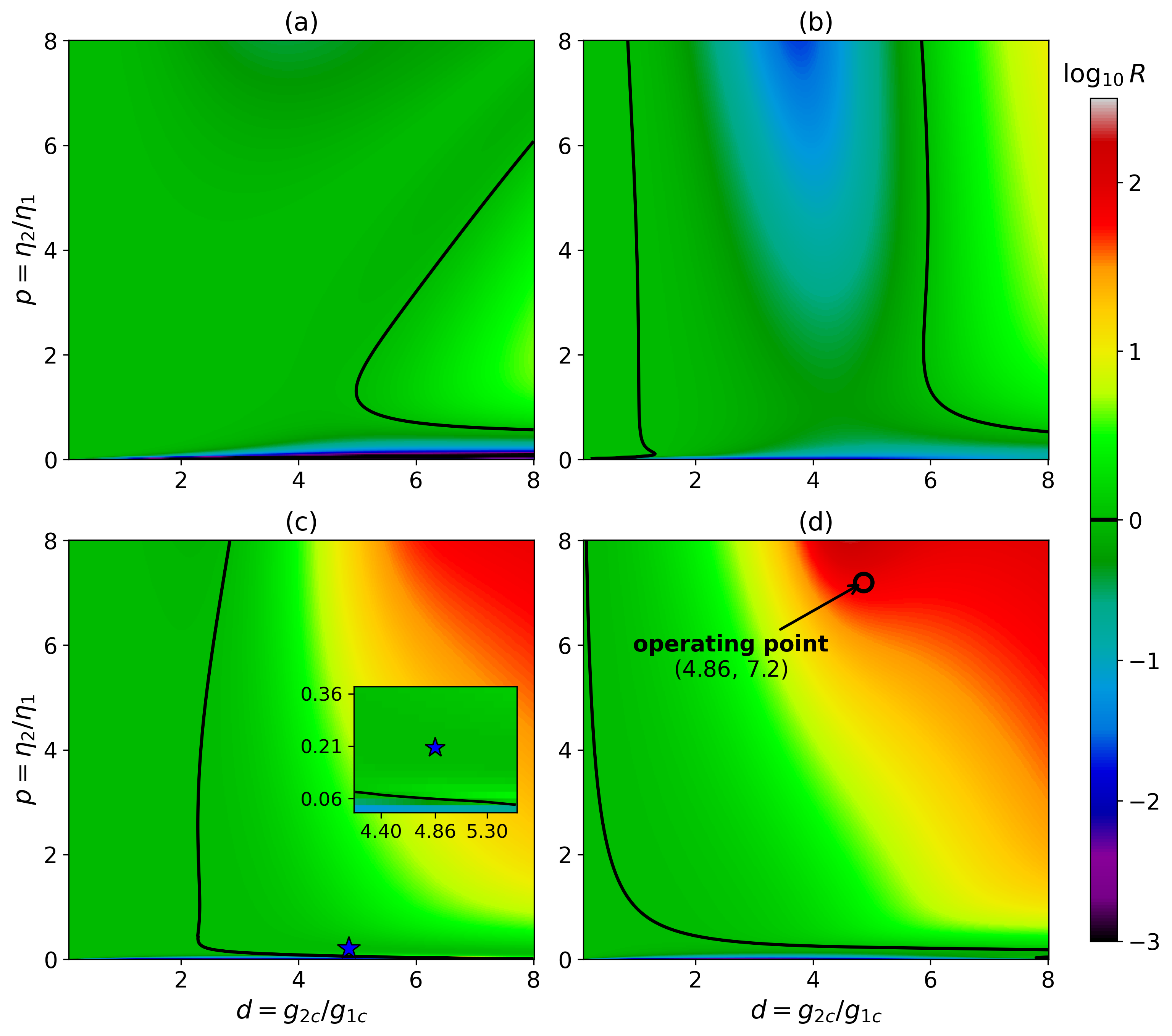}
\caption{Two-dimensional colour maps of the CSI violation factor $R$, on a
$\log_{10}$ scale as functions of the coupling ratio $d=g_{2c}/g_{1c}$ and the
drive ratio $p=\eta_2/\eta_1$ at (a) $\phi=0$, (b) $\phi=\pi/2$,
(c) $\phi=\pi$ and (d) the operating phase $\phi^{*}=1.183\pi$.  The
white curve is the $R=1$ contour separating the classical region from the
CSI-violating one and the black line on the colour bar marks
$\log_{10}R=0$. Also, the white circle in (d) marks the operating point $(d,p)=(4.86,7.2)$ where $R=73.4$.  The white star in (c) and the region enlarged in the inset mark the classical reference point $(d,p)=(4.86,0.206)$ at which $R=1$ at $\phi=\pi$.  All the system parameters are taken from Refs.~\cite{MagnonBlockade_2,nair2022cavity,fan2024quantum,SuperconductingQubit_2,DirectCouplingBetweenMagnonandQubit_2}:
$\gamma_1=\gamma_2=1.3/(2\pi)$~MHz, $\Gamma_q=1.74/(2\pi)$~MHz,
$g_1=15/(2\pi)$~MHz, $\eta_1=0.001/(2\pi)$~MHz and $g_q=189/(2\pi)$~MHz.}
\label{fig:2}
\end{figure}

In Fig. 2, we examine the CSI violation factor $R$ over the coupling ratio and
drive-ratio at $0.1\le d\le8$ and $0.01\le p\le8$ respectively. The cavity drive detuning is held fixed at $\Delta_c=5g_q$ so that the adiabatically eliminated Hamiltonian of Eq.~(3) is valid uniformly across the entire plane.  As established in Appendix~A, $R>1$ signifies the nonclassical correlations that cannot be reproduced by any classical mixture of coherent states while $R\le1$ indicates their
absence~\cite{PhysRevLett.110.267004}.

The relative drive phase controls the extent of the violating region in which $R>1$ grows from $15.3\%$ at $\phi=0$ through $35.2\%$ at $\phi=\pi/2$ and $69.8\%$ at $\phi=\pi$ to $92.6\%$ at the operating phase $\phi^{*}=1.183\pi$. Therefore, a system moves across the classical bound at a fixed coupling and drive ratio by adjusting $\phi$ alone.  At the operating point $(d,p)=(4.86,7.2)$ the violation factor takes the values $R=0.564$, $0.151$, $12.22$ and $73.4$ at $\phi=0,\pi/2,\pi$ and $\phi^{*}$ respectively. 

We deliberately do not reported the largest value of $R$ in each panel because $R$ diverges at the blockade conditions $C_{g20}=0$ and $C_{g02}=0$ so any near-singular value is limited by resolution rather than physical. Moreover, the grid maxima lie on the $d=8$ or $p=8$ boundaries within our observed ranges which is only reflecting the extent of the scan rather than interior features. So, a meaningful figure of merit must instead be evaluated at a non-singular, pair-correlated operating point. We therefore adopt the criteria $0.02<g_{11}^{(2)}(0),g_{22}^{(2)}(0)<0.6$ and $g_{12}^{(2)}(0)>1.5$ respectively.

Subject to these criteria, the operating point is an interior optimum in $d$. At fixed value of $p=7.2$ no non-singular pair-correlated point exists for $d\lesssim4.7$ or $d\gtrsim7$ while within this window the phase-optimised violation decreases monotonically with $d$ where we observe that $R=73.4$ at $d=4.86$, $68.8$ at $d=5.50$, and $65.7$ at $d=6.50$. The optimal phase drifts correspondingly from $\phi^{*}/\pi=1.183$ at $d=4.86$ to $1.156$ at $d=5.50$, and $1.135$ at $d=6.50$ respectively. Also, at the operating point both modes are individually antibunched while the pair correlation is strongly super-Poissonian $g_{11}^{(2)}(0)=0.148$, $g_{22}^{(2)}(0)=0.568$, and $g_{12}^{(2)}(0)=2.486$. This simultaneous suppression of two-magnon excitations with enhanced joint excitation is precisely the configuration that violates the classical inequality.

The enhancement arises from the two-pathway interference mechanism established in Eqs.~(A13)--(A18)(Appendix A). Here, the destructive interference suppresses the same-mode two-magnon pathways to $|g,2,0\rangle$ and $|g,0,2\rangle$ while the cross-mode pathway to $|g,1,1\rangle$ is simultaneously enhanced by constructive interference. The combination of both effects produces the strong violation since $R$ depends on the ratio of the enhanced cross-mode amplitude to the suppressed same-mode amplitudes as shown in Eq.~(A23). The inset in Fig. 2(c) shows that the same-mode amplitudes $C_{g20}$ and $C_{g02}$ are both suppressed while $C_{g11}$ remains at this typical weak driving regime so that $R$ is exactly at the classical bound which is indeed approached from the near-coherent side rather than from the violating side (see Fig.~4). Again at the same $(d,p)$, $R$ falls below unity which are $0.069$, $0.198$, and $0.887$ at $\phi=0$, $\pi/2$, and $\phi^{*}$, respectively. This demonstrates that the classical bound is reached at weak driving regardless of phase and that the strong violation is a strong driving phenomenon as well.

The solid white $R=1$ contour in each figure is determined numerically from the exact steady-state solution and its shape admits no simple closed form expression appears once the qubit-mediated dressing of Eqs.~(A16)--(A18) is included (see Appendix~A). These results establish the relative drive phase $\phi$ as a continuously tunable control parameter for fixed $(d,p)$ by adjusting $\phi$ alone moves the system between $R\ll1$ and $R\gg1$ which offers a practical route for switching and scaling quantum correlations in magnon-based quantum information architectures.


\begin{figure}[htbp]
\centering
\includegraphics[width=0.48\textwidth]{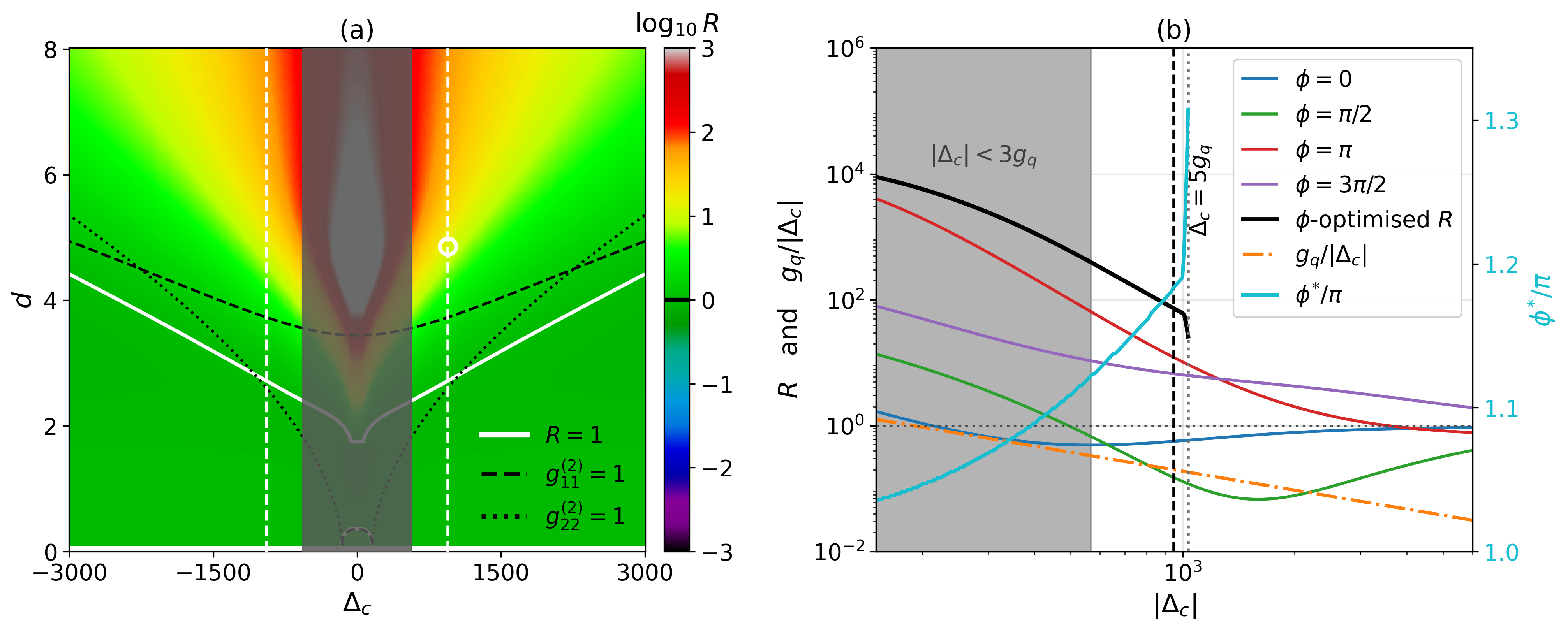}
\caption{Cavity-detuning dependence of the two-mode magnon blockade and the CSI violation at the drive ratio $p=\eta_2/\eta_1=7.2$.  
(a) Violation factor $R$ on a $\log_{10}$ scale versus the cavity--drive detuning $\Delta_c$ and the coupling ratio $d=g_{2c}/J_{1c}$ at fixed $\phi=\pi$.
Three contours partition in the plane: the classical bound $R=1$ (white solid) and the antibunching boundaries $g_{11}^{(2)}(0)=1$ (black dashed) and $g_{22}^{(2)}(0)=1$ (black dotted) above which the respective Kittel mode is antibunched.  
The dark band marks $|\Delta_c|<3g_q$ where the dispersive
Hamiltonian of Eq.~(3) is not reliably valid and the white dashed lines mark the operating detuning $|\Delta_c|=5g_q=945$ and the circle is the operating point $(945,4.86)$.  
(b) $R$ versus $|\Delta_c|$ at $d=4.86$, $p=7.2$ for
the four relative phases $\phi=0,\pi/2,\pi$ and $3\pi/2$. The
phase-optimised $R$ restricted to non-singular pair-correlated operating points (black) and the dispersive expansion parameter $g_q/|\Delta_c|$ (orange, dash-dotted). Also, the right axis (cyan) gives the optimal phase $\phi^{*}$.  The optimised curves terminate at the vertical dotted line which is at $|\Delta_c|\simeq1.04\times10^{3}$ ($5.5g_q$) beyond which the second mode is no longer antibunched.  All values follow from the exact solution of
Eqs.~(A2)--(A9).}
\label{fig:detuning_map}
\end{figure}

Fig. 3 establishes the detuning range where Eq.~(3) remains valid and quantifies the violation cost of operating deeper within it. Since all derived parameters scale as $1/\Delta_c$ we exclude $|\Delta_c|<3g_q$ and operate at $|\Delta_c|=5g_q$ where the expansion parameter is $g_q/|\Delta_c|=0.200$. The color map in Fig.~5(a) is symmetric about $\Delta_c=0$ which arises from conjugating Eqs.~(A2)--(A9) whose collapse operators are real and the first drive has $\phi_1=0$ which gives the relation $R(\Delta_c,d,\phi)=R(-\Delta_c,d,-\phi)$ that reduces to origin symmetry at $\phi=\pi$. We have verified this numerically to zero deviation and hence we discuss only $|\Delta_c|$.

The violation is extensive in $R>1$ which is over $60.8\%$ of the scanned plane. The three contours demarcate the regimes of interest and their relative ordering changes across the plot. As $d$ increases, the qubit-mediated couplings $J_2=d g_1 g_q/\Delta_c$ and $J_{12}=d g_1^2/\Delta_c$ strengthen so that each mode becomes antibunched and the correlations become nonclassical only above its contour. The useful region lies above all three curves where both Kittel modes are individually antibunched and violate the classical Cauchy-Schwarz inequality (CSI). Its lower edge is set by whichever threshold is highest and the threshold that binds is determined by how each mode acquires its blockade. The second mode is driven and coupled directly so it becomes blockaded at a modest coupling ratio. In contrast, the first mode is blockaded only through the cross-coupling $J_{12}$ which is weaker than $J_2$ by the fixed factor $g_1 / g_q \simeq 0.08$. The first mode therefore requires a substantially larger $d$ and the condition $g_{11}^{(2)}(0) = 1$ constitutes the binding constraint throughout most of the figure. At $|\Delta_c| = 700$, mode~2 is antibunched above $d = 2.12$ and the classical bound is crossed above $d = 2.51$. However, the useful region begins only at $d = 3.61$ where mode~1 antibunches. The two antibunching boundaries cross near $(|\Delta_c|, d) \approx (2.2 \times 10^{3}, 4.5)$. So the couplings have weakened sufficiently that the mode~2 becomes the limiting one instead and its threshold sets the lower edge beyond that detuning.

A further feature is worth noting at $|\Delta_c| \gtrsim 1.2 \times 10^{3}$ where we see that the $R = 1$ contour lies below both antibunching boundaries so the correlations turn nonclassical before either mode is antibunched. In this range the CSI is the least restrictive of the three conditions and it is the simultaneous antibunching of both modes rather than the violation itself which is limiting the accessible operating region. The operating point $(|\Delta_c|, d) = (945, 4.86)$ lies above all three contours yielding $R = 12.2$ with $g_{11}^{(2)}(0) = 0.589$, $g_{22}^{(2)}(0) = 0.603$, and $g_{12}^{(2)}(0) = 2.08$ at $\phi = \pi$.

Fig. 3(b) makes the trade-off more explicit. Here, $\phi=\pi$ gives the largest violation among the four phases at the operating detuning where $R=12.2$ against $0.56$, $0.15$ and $6.69$ at $\phi=0$, $\pi/2$ and $3\pi/2$ respectively. 
$R$ falls from $2451$ in $|\Delta_c|=300$ to $395$ in $3g_q$ and $73.4$ in the chosen optimum $5g_q$ because $J_1,J_2,J_{12}$ and
$\Delta_1,\Delta_2,\Delta_g$ all shrink as $1/\Delta_c$ while dissipative rates remain fixed. 
So the system is driven towards the linear coherent limit in which $R\to1$ at the same rate as the dispersive approximation improves which is exactly given by Eq.~(A24). 

The same mechanism ends the optimised curves.  As the couplings weaken both autocorrelations rise towards the coherent value $g^{(2)}(0)=1$ and the strongly driven second mode loses its blockade first which passes
$g_{22}^{(2)}(0)=0.6$ at $|\Delta_c|\simeq5.5g_q$.  The two modes cannot be antibunched and pair-correlated simultaneously beyond that point so the optimisation has no admissible solution and hence the termination of the curve shows a physical boundary rather than a limitation of the scan.

Finally, the optimal phase drifts systematically with detuning from $\phi^{*}/\pi=1.069$ at $|\Delta_c|=300$
through $1.122$ at $3g_q$ to $1.183$ at $5g_q$. Since $R\equiv1$ at every
phase when $J_1=J_2=0$ [Eq.~(A24)] so any phase dependence originates entirely in the qubit-mediated terms whose relative weight varies with $\Delta_c$. The consequences for the choice of operating phase are examined in next Figure.

%

\begin{figure*}[htbp]
\centering
\includegraphics[width=\textwidth]{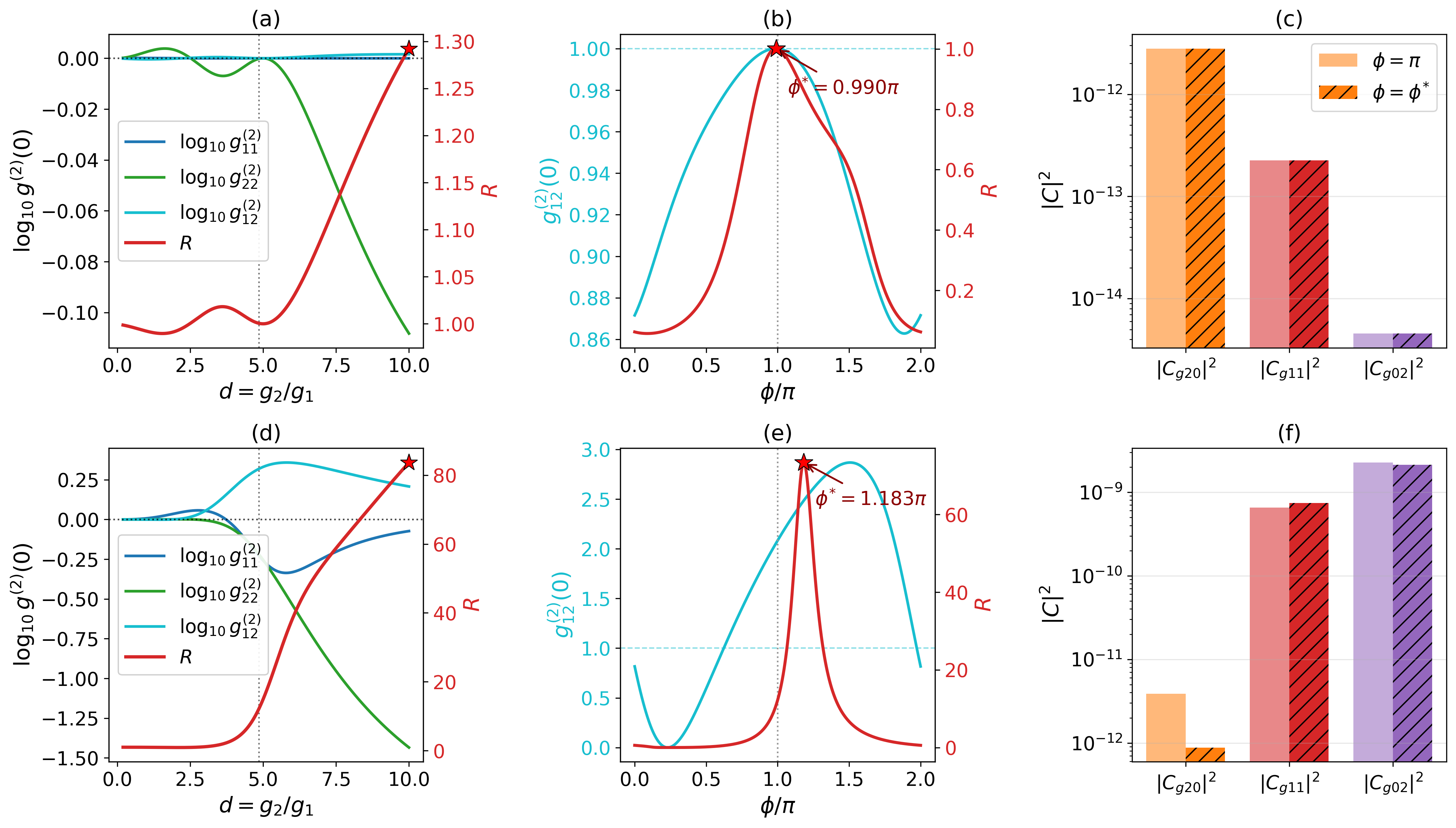}
\caption{Top row, $p=\eta_2/\eta_1=0.2$; bottom row, $p=7.2$.
(a),(d) The normalised correlations $g_{11}^{(2)}(0)$,
$g_{22}^{(2)}(0)$, $g_{12}^{(2)}(0)$ (left axis, $\log_{10}$) and the
violation factor $R$ (right axis) as functions of the coupling ratio $d$, at
fixed relative phase $\phi=\pi$; the grey dotted line marks the operating
value $d=4.86$ and the star marks the largest $R$ within the scanned range
$0.2\le d\le10$. (b),(e) $g_{12}^{(2)}(0)$ (left) and $R$ (right)
as functions of the relative phase $\phi$ at $d=4.86$; the grey dotted line
marks the control phase $\phi=\pi$, and in (e) the star marks the
numerically located maximum at $\phi^{*}=1.183\pi$.  (c),(f) The
two-magnon amplitudes $|C_{g20}|^2$, $|C_{g11}|^2$, $|C_{g02}|^2$ at the
operating coupling ratio which are evaluated at the control phase $\phi=\pi$ (solid)
and at $\phi^{*}$ (hatched) \cite{jin2023magnon,PhysRevA.98.013821}.}
\label{fig:weak_strong}
\end{figure*}

In figure 4, we study the near classical and strongly correlated regimes by
comparing two drive ratios at $p=0.2$ and $p=7.2$ which are at the operating coupling
ratio $d=4.86$ and detuning $\Delta_c=5g_q$ respectively. We emphasise that both regimes
lie within the weak driving limit in absolute terms. The single-magnon
occupations remain far below unity throughout the limit so that $p$ is the ratio of
two small drives but not a measure of driving strength. Also, the transition described
here is controlled by the interference between the two drives which is set by their
ratio and relative phase and not by any increase in total driving power.

At $p=0.2$ the system is essentially classical.  As shown in Fig. 4(a), the
three correlation functions remain at $g_{11}^{(2)}(0)\approx
g_{22}^{(2)}(0)\approx g_{12}^{(2)}(0)\approx1$ across the entire
coupling ratio range and $R$ stays within a fraction of a percent of the
classical bound so that it reaches only $R\approx1.29$ at the very edge of the scanned
range.  At the operating point $d=4.86$ and $\phi=\pi$, we obtain the values 
$g_{11}^{(2)}(0)=1.000$, $g_{22}^{(2)}(0)=1.000$, $g_{12}^{(2)}(0)=1.000$ and
$R=1.0005$. Also, the cross-correlation factorises as
$g_{12}^{(2)}(0)\approx[g_{11}^{(2)}(0)\,g_{22}^{(2)}(0)]^{1/2}$ so that the
condition of near-coherent two-mode statistics for which the Cauchy-Schwarz
bound is saturated but not violated.  In Fig. 4(b), we confirms that $R$ and
$g_{12}^{(2)}(0)$ remain at or below unity across the full phase range at this
drive ratio and the phase modulation of $R$ is smaller than $10^{-3}$. Since the system has not produced the interference necessary to deviate from Gaussian-like statistics so no optimal phase is established. Therefore, the
corresponding two-magnon amplitudes, as shown in Fig. 4(c), are of the order of $10^{-13}$ and
can only be weakly redistributed between $\phi=\pi$ and $\phi^{*}$ as
expected when there is no blockade active at all.

At $p=7.2$ the behaviour changes qualitatively when we study the coupling ratio
at fixed $\phi=\pi$, as shown in Fig. 4(d). We observe that $\log_{10}g_{11}^{(2)}(0)$ crosses zero near
$d\approx3.7$ beyond which mode 1 becomes antibunched while $R$ rises
steadily with $d$.  At the operating point the control phase $\phi=\pi$
already yields $R=12.2$ with $g_{11}^{(2)}(0)=0.589$, $g_{22}^{(2)}(0)=0.603$
and $g_{12}^{(2)}(0)=2.08$.  Fig. 4(e) shows that tuning the relative phase
away from $\pi$ sharply increases the violation and $R$ rises from $12.2$ at
$\phi=\pi$ to a maximum $R=73.4$ at $\phi^{*}=1.183\pi$, where
$g_{11}^{(2)}(0)=0.148$, $g_{22}^{(2)}(0)=0.568$ and $g_{12}^{(2)}(0)=2.49$.
Both modes are individually antibunched while the pair correlation is
strongly super-Poissonian. This confirms that the configuration violates the classical
inequality. The two modes are emitted as correlated magnon pairs while
same-mode double excitation is suppressed.  This maximum is an interior
feature of the phase scan which is offset from the control phase $\phi=\pi$ by
$+0.183\pi$. It is not predicted by any qubit-decoupled phase condition
since in the $J_1=J_2=0$ limit $R$ is identically unity at every phase
[Eq.(A24)] and it is therefore located numerically.

The microscopic picture is given by the amplitude bars, as shown in Fig. 4(f), and it is more asymmetric by a straightforward description that suppresses both same-mode states.  At the control phase $\phi=\pi$ the three two-magnon
amplitudes are strongly ordered as $|C_{g02}|^2=2.26\times10^{-9}$,
$|C_{g11}|^2=6.53\times10^{-10}$ and $|C_{g20}|^2=3.87\times10^{-12}$. Here, the
same-mode amplitude of the strongly driven mode $|C_{g02}|^2$ is in fact
the largest of the three while the same-mode amplitude of the weakly
driven mode $|C_{g20}|^2$ lies more than two orders of magnitude below the cross-mode amplitude $|C_{g11}|^2$. Therefore, the two
same-mode channels are not suppressed together. Here, only mode 1
is blockaded while mode 2 remains strongly populated.  This asymmetry is a
direct consequence of the coupling ratio $d=g_{2c}/g_{1c}=4.86$ where mode~2 couples to
the qubit roughly five times more strongly than mode~1 which is driven
correspondingly harder and stays populated at the two-magnon level whereas
mode~1 is suppressed.  It is consistent with Fig.~4(d) where
$g_{11}^{(2)}(0)$ falls well below unity while $g_{22}^{(2)}(0)$ remains near
$0.6$.

The reason that the single-mode blockade is sufficient to drive a large violation
lies in the structure of $R$. In Eq.~(A23),
$R=|C_{g11}|^4/(4|C_{g20}|^2|C_{g02}|^2)$ depends on the
product $|C_{g20}|^2|C_{g02}|^2$ but not on either same-mode
amplitude individually so a strong suppression of one same-mode channel drives
$R\gg1$ even while the other same-mode channel remains large. The tiny
$|C_{g20}|^2$ in the denominator is not offset by the large $|C_{g02}|^2$
because their product is what makes it meaningful.  Moving from the control phase to optimal
$\phi^{*}=1.183\pi$ deepens precisely this suppression. $|C_{g20}|^2$
falls further from $3.87\times10^{-12}$ to $8.82\times10^{-13}$ by a
factor of $4.4$ so that the ratio $|C_{g11}|^2/|C_{g20}|^2$ grows from $\approx170$
to $\approx840$ while $|C_{g11}|^2$ and $|C_{g02}|^2$ are left almost
unchanged.  The violation factor rises accordingly from $R=12.2$ to $R=73.4$.
This selective deepening of the mode-1 blockade by the relative drive phase
at fixed coupling and drive ratios is the mechanism by which the system is
tuned from weak to strong Cauchy-Schwarz violation.

%

\begin{figure*}[htbp]
\centering
\includegraphics[width=\textwidth]{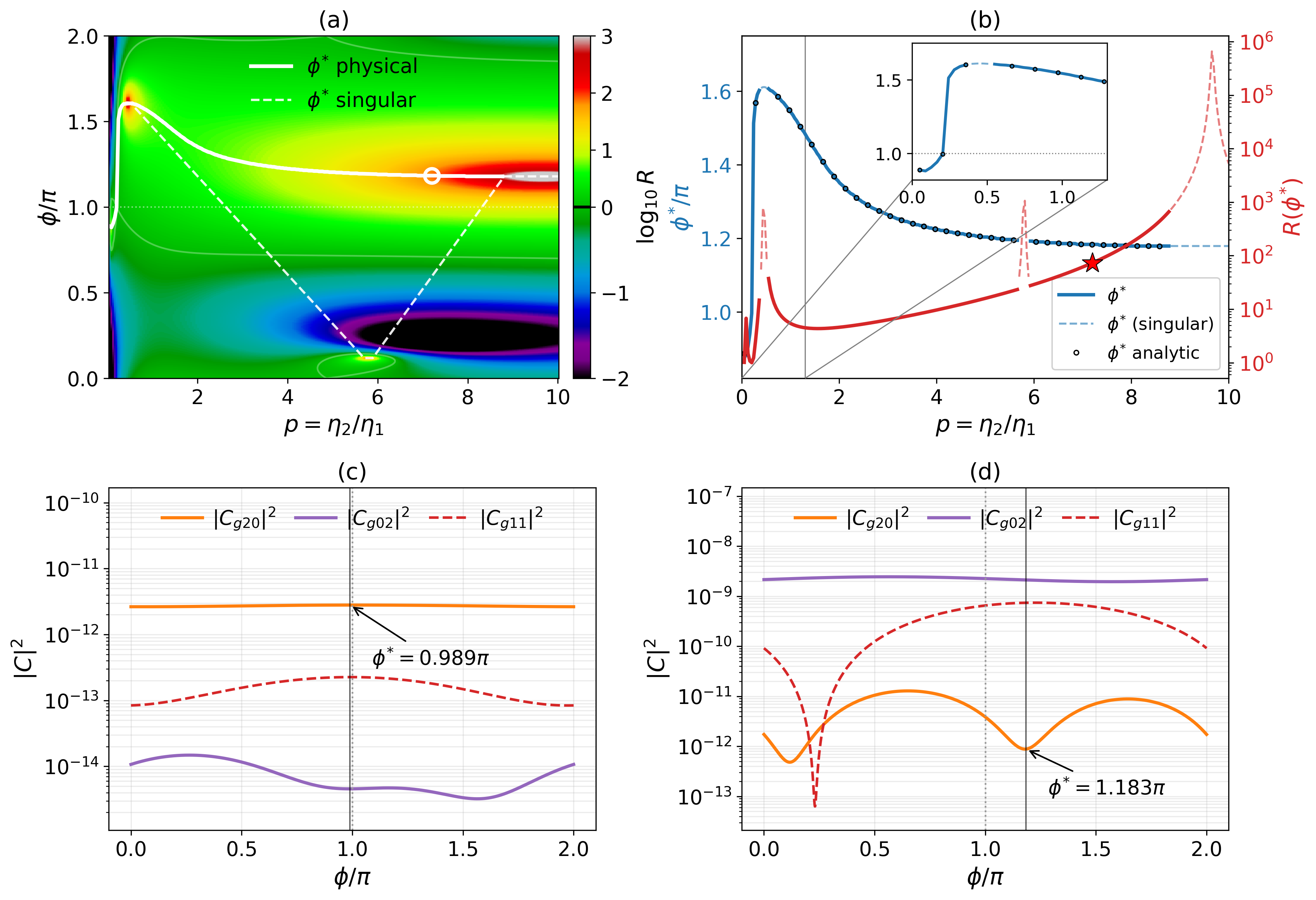}
\caption{Phase structure of the CSI violation as a function of the drive
ratio $p=\eta_2/\eta_1$ at the operating coupling ratio $d=4.86$ and
detuning $\Delta_c=5g_q$.  (a) Colour map of $R(p,\phi)$ on a
$\log_{10}$ scale (the classical bound $\log_{10}R=0$ is marked on the colour
bar).  The white curve is the numerically traced optimal phase
$\phi^{*}(p)$ which is drawn solid where the optimum is a non-singular,
pair-correlated operating point and dashed where it degenerates into a
blockade spike. The circle marks the operating point at $p=7.2$.
(b) The traced optimal phase $\phi^{*}$ (blue, left axis) and the
corresponding violation factor $R(\phi^{*})$ (red, right axis, log scale)
versus $p$ where the solid and dashed segments follow the same physical and singular
convention. Also, the star marks the
operating value $R=73.1$ at $p=7.2$.  The inset enlarges the low-drive region
$p\le1.3$ where $\phi^{*}$ switches branches.  (c),(d) The three
two-magnon amplitudes $|C_{g20}|^2$ (mode 1), $|C_{g02}|^2$ (mode 2) and
$|C_{g11}|^2$ (cross) as functions of the relative phase $\phi$ at
$p=0.2$ (c) and $p=7.2$ (d); the grey dotted line marks $\phi=\pi$ and the
black line marks the numerically located optimum $\phi^{*}$.}
\label{fig:phase_structure}
\end{figure*}

Fig.5 traces how the optimal relative drive phase
and the accessible violation evolve as the drive ratio is increased at the
operating coupling ratio $d=4.86$ and detuning $\Delta_c=5g_q$.  Throughout
this figure the optimal phase $\phi^{*}(p)$ is obtained directly from the
exact steady-state solution as we have shown that no closed-form prediction is
available in this regime.

Fig. 5(a) shows the full $(p,\phi)$ map where a single connected ridge of large
$R$ runs across the plane and the traced optimum $\phi^{*}(p)$ follows it.
For $p\gtrsim1$, the ridge is a genuine operating region at which both
same-mode states are suppressed but remain non-singular and the two modes are
positively pair-correlated. The optimal phase drifts smoothly downward
with increasing drive from $\phi^{*}=1.54\pi$ at $p=1$ through $1.22\pi$
at $p=4$ to $1.18\pi$ at the operating point $p=7.2$.  Fig. 5(b) makes this
quantitative which shows that along the physical branch $R(\phi^{*})$ rises monotonically
from $R\approx5$ at $p=1$ to $R=29$ at $p=6$ and $R=73.1$ at $p=7.2$.  The
violation is therefore a strong-drive ratio phenomenon that switches on
continuously and the operating phase must be re tuned as $p$ is changed.

There are two aspects of the traces that we discuss now.  First, at low drive the optimal
phase switches branches abruptly. As $p$ increases through
$p\approx0.2$--$0.28$, $\phi^{*}$ jumps from just below $\pi$ to
$\approx1.6\pi$ which is shown enlarged in the inset of Fig. 5(b).  This is the point
at which the weak-driving maximum near $\phi=\pi$ where $R$ merely
approaches the classical bound from below which gives way to the genuine violating
branch. There is no violation to optimise below the switch and the reported
$\phi^{*}$ simply tracks the shallow weak-drive maximum.  Second, the
dashed segments of both traces shows the isolated feature near $p\approx0.5$
and the tail beyond $p\approx8.5$. We obtain values of $p$ at which the maximum
of $R$ over $\phi$ occurs at a blockade zero where a same-mode amplitude
$C_{g20}$ or $C_{g02}$ is driven almost exactly to zero
 that is $g_{11}^{(2)}(0)$ or $g_{22}^{(2)}(0)<0.02$.  The apparent values obtauined as 
$R=1233$ at $p=9$ and $R=5261$ at $p=10$ are set by how close the scan lands
to the zero rather than by a robust operating condition and the values are excluded
from the physical interpretation. We retain them only to show that the
physical ridge and the singular spikes are continuously connected.

Fig. 5(c) and 5(d) resolve the microscopic mechanism by displaying the three
two-magnon amplitudes as functions of $\phi$.  At $p=0.2$, all
three amplitudes are essentially independent of phase as shown in Fig.5(c). $|C_{g20}|^2$ does
not vary at all over the scan and $|C_{g02}|^2$ varies by less than a factor
of five so that no phase can produce a blockade and $R$ stays at the classical
bound. This is the weak-drive regime in which the system has not developed the
interference required to depart from Gaussian-like statistics.  At $p=7.2$,
Fig. 5(d), the results are entirely different and strongly asymmetric between
the two modes. Here, the same mode amplitude $|C_{g20}|^2$ of mode-1 swings by a
factor of $27$ across the phase and passes through a sharp minimum while the
mode-2 amplitude $|C_{g02}|^2$ barely moves and
remains the largest of the three.  The optimum $\phi^{*}=1.183\pi$ sits at
the minimum of $|C_{g20}|^2$. So it is the phase at which the weakly driven mode
is most deeply blockaded  since $R=|C_{g11}|^4/(4|C_{g20}|^2|C_{g02}|^2)$ in
Eq.(A23) depends on the product $|C_{g20}|^2|C_{g02}|^2$. This
single-mode suppression is sufficient to drive $R\gg1$ even though the other
same-mode channel stays large which is consistent with the amplitude ordering
discussed for Fig. 4(f).

Finally, we show that the optimal phase is captured analytically as well. The relative phase enters Eqs. (A2)-- (A9) only through the combination
$\eta_2 e^{-i\phi}$ and at most quadratically along any pathway to a
two-magnon state where each two-magnon amplitude is an exact quadratic in
$z=e^{-i\phi}$ [Eq. (A25)] and $R$ depends on $\phi$ through the
first and second harmonic forms. Thus the optimal phase $\phi^{*}(p)$ is the relevant root of the single closed form condition in Eq. (A26) which is obtained by imposing a stationary state of the violation factor $\partial_{\phi}\ln R=0$.  The open circles in
Fig.5(b) show this analytic aspects clearly. It
coincides with the numerically determined optimum across the entire physical
branch from $\phi^{*}=1.54\pi$ at $p=1$ to $1.18\pi$ at the operating
point and reproduces the corresponding violation factor to better than one
part in $10^{3}$.  
In practice the optimum remains close to $\phi=\pi$ and hence the offset
$\phi^{*}-\pi$ decreases monotonically from $+0.54\pi$ at $p=1$ to
$+0.18\pi$ at $p=7.2$ so that $\phi=\pi$ provides a convenient starting
point with the residual offset fixed by Eq. (A26) or tuned in situ.

\section{Conclusion}

In summary, we have shown that two Kittel magnon modes coupled through a common
superconducting qubit can be driven into a strongly nonclassical state that
violates the Cauchy-Schwarz inequality. The qubit induces both a magnon--magnon coupling and mode--qubit
couplings after adiabatically eliminating the
cavity and solving the steady state in the two-magnon manifold reveals the
underlying mechanism. The relative drive phase tunes an interference that
suppresses the same-mode amplitudes and antibunches each mode while enhancing
the cross-mode amplitude so that at the optimal phase $\phi^{*}\simeq1.18\pi$
the violation factor becomes large with $R\gg1$. The effect is intrinsically
quantum and appears only under strong driving since without the qubit the
system is Gaussian and cannot violate the classical bound and the violation
vanishes in the weak-drive limit. We verified the analysis against the exact
numerics and confirmed the accuracy of the cavity elimination and we found the
violation to be robust against realistic magnon and qubit dissipation. These
results establish cavity magnonics with qubit as a practical and
phase-controlled route to on-demand nonclassical magnon correlations that can
be extended naturally toward full counting statistics, inter-mode entanglement,
and finite-temperature operation.

\begin{acknowledgments}
S. A. B. would like to acknowledge Indian Institute of Technology, Guwahati for providing financial support for the research.
\end{acknowledgments}

\appendix

\section{Full Analytical Derivations}

To provide physical insight into the statistical properties of magnons and the violation of the Cauchy-Schwarz Inequality (CSI), we establish an analytical approach using the Schr\"odinger equation in the weak-driving limit.

\subsection{State Truncation and Wave Function}

We truncate the Hilbert space to the two-magnon excitation subspace under weak driving so that the low energy levels are significantly populated.
The wave function $|\psi\rangle$  of the system is expanded as
\begin{flalign}
&|\psi\rangle \approx \; C_{g00}|g,0,0\rangle + \underbrace{C_{g10}|g,1,0\rangle + C_{g01}|g,0,1\rangle}_{\text{1-magnon subspace}} \notag&\\
&\quad + \underbrace{C_{g11}|g,1,1\rangle + C_{g20}|g,2,0\rangle + C_{g02}|g,0,2\rangle}_{\text{2-magnon subspace}} \notag&\\
&\quad + \underbrace{C_{e00}|e,0,0\rangle + C_{e10}|e,1,0\rangle + C_{e01}|e,0,1\rangle}_{\text{qubit excitation subspace}}
\tag{A1}&
\end{flalign}
Here, $|g\rangle$ and $|e\rangle$ are the qubit ground and excited states
while $|m_1,m_2\rangle$ denotes the Fock states of the two Kittel modes.
$|C_{jm_1m_2}|^2$ is the occupation probability of the corresponding state.

\subsection{Dynamical Evolution}

Non-Hermitian decay is incorporated by taking the detunings to be complex quantities $\Delta_1 \equiv \tilde\Delta_{m_1} - i\gamma_1/2$, $\Delta_2 \equiv \tilde\Delta_{m_2} - i\gamma_2/2$, and $\Delta_g \equiv \tilde\Delta_q - i\Gamma_q/2$. The qubit--magnon couplings are denoted by $J_1\equiv g_{1q}$ and $J_2\equiv g_{2q}$. Therefore using the Schr\"odinger equation $i\,\partial_t|\psi\rangle = \hat H_{\mathrm{non-Herm}}|\psi\rangle$ onto each basis state in Eq.~(A1) and imposing the steady-state condition $\dot C_n = 0$ gives the following coupled equations for the one-magnon manifold.
\begin{flalign}
&\Delta_1 C_{g10} + J_{12}C_{g01} + J_1 C_{e00} + \eta_1 C_{g00} = 0 \tag{A2}&\\
&\Delta_2 C_{g01} + J_2 C_{e00} + J_{12}C_{g10} + \eta_2 e^{-i\phi} C_{g00} = 0 \tag{A3}&\\
&\Delta_g C_{e00} + J_1 C_{g10} + J_2 C_{g01} = 0 \tag{A4}&
\end{flalign}
Also, the two-magnon manifold is given by
\begin{flalign}
&2\Delta_1 C_{g20} + \sqrt2 J_1 C_{e10} + \sqrt2 J_{12} C_{g11} + \sqrt2\eta_1 C_{g10} = 0 \tag{A5}&\\
&(\Delta_1+\Delta_2) C_{g11} + J_1 C_{e01} + J_2 C_{e10} + \sqrt2 J_{12}(C_{g20}+C_{g02}) \notag&\\
&\quad + \eta_1 C_{g01} + \eta_2 e^{-i\phi} C_{g10} = 0 \tag{A6}&\\
&2\Delta_2 C_{g02} + \sqrt2 J_2 C_{e01} + \sqrt2 J_{12} C_{g11} \notag&\\ & \quad+ \sqrt2\eta_2 e^{-i\phi} C_{g01} = 0 \tag{A7}&\\
&(\Delta_1+\Delta_g) C_{e10} + \sqrt2 J_1 C_{g20} + J_2 C_{g11}\notag&\\
&\quad + J_{12} C_{e01} + \eta_1 C_{e00} = 0 \tag{A8}&\\
&(\Delta_2+\Delta_g) C_{e01} + \sqrt2 J_2 C_{g02} + J_1 C_{g11} \notag&\\ & \quad + J_{12} C_{e10} + \eta_2 e^{-i\phi} C_{e00} = 0 \tag{A9}&
\end{flalign}
These equations follow directly from the ladder-operator matrix elements of
$\hat H_{\mathrm{non-Herm}}$ in the truncated basis (A1). Here, \
$J_{12}(m_1^\dagger m_2+m_2^\dagger m_1)$ connects
$|g,2,0\rangle\leftrightarrow|g,1,1\rangle$ and
$|g,1,1\rangle\leftrightarrow|g,0,2\rangle$ with amplitude $\sqrt2 J_{12}$ in
each direction and $J_2(m_2^\dagger\sigma_-+m_2\sigma_+)$ connects
$|e,1,0\rangle\leftrightarrow|g,1,1\rangle$ with unit amplitude.

\subsection{One-Magnon Amplitudes}

Solving Eqs.~(A2)--(A4) exactly for the single-magnon amplitudes yields
\begin{flalign}
&C_{g10} = \frac{-\eta_1(\Delta_2\Delta_g - J_2^2) + \eta_2 e^{-i\phi}(J_{12}\Delta_g - J_1J_2)}{\mathcal D_1}
\tag{A10}&
\end{flalign}
\begin{flalign}
&C_{g01} = \frac{-\eta_2 e^{-i\phi}(\Delta_1\Delta_g - J_1^2) + \eta_1(J_{12}\Delta_g - J_1J_2)}{\mathcal D_1}
\tag{A11}&
\end{flalign}
\begin{flalign}
&\mathcal D_1 = (\Delta_1\Delta_2 - J_{12}^2)\Delta_g - \Delta_1 J_2^2 - \Delta_2 J_1^2 + 2J_1J_2J_{12}
\tag{A12}&
\end{flalign}

\subsection{Two-Magnon Amplitudes: Perturbative Solution Including the Qubit}
\label{app:twomagnon}

A closed $5\times5$ linear system for $C_{g20},C_{g11},C_{g02},C_{e10}$ , and $C_{e01}$ are formed by all the equations from (A5)--(A9) along with (A10)--(A12) substituted as sources. Note that the system lacks a simple closed form once $J_1,J_2,J_{12}$ are kept broad. We instead solve it perturbatively in the coupling strengths parametrized as $J_1,J_2,J_{12}=O(\varepsilon)$ with $\varepsilon\to1$ at the end which is consistent with the weak qubit--magnon coupling regime. Each amplitude is written as  $X = X^{(0)} + X^{(1)} + X^{(2)} + O(\varepsilon^3)$ and the recursion matrix is 
$M_0\,\mathbf{x}^{(n)} = -\big(\mathbf S^{(n)} + M_1\,\mathbf{x}^{(n-1)}\big)$ (where $M_0=\mathrm{diag}(2\Delta_1,\ \Delta_1{+}\Delta_2,\ 2\Delta_2,\Delta_1{+}\Delta_g,\ \Delta_2{+}\Delta_g)$ is the bare part and $M_1$ is the
$O(\varepsilon)$ coupling matrix) and it will be solved order by order. 

The uncoupled zeroth order occupation in the absence of $J_1, J_2$ and $J_{12}$ is given by 
\begin{flalign}
&C_{g20}^{(0)} = \frac{\eta_1^2}{\sqrt2\,\Delta_1^2}, \qquad
C_{g11}^{(0)} = \frac{\eta_1\eta_2 e^{-i\phi}}{\Delta_1\Delta_2} = C_{g10}^{(0)}C_{g01}^{(0)}, \notag\\
&C_{g02}^{(0)} = \frac{\eta_2^2 e^{-2i\phi}}{\sqrt2\,\Delta_2^2}.
\tag{A13}&
\end{flalign}

This is simply the product state of two independently driven modes and it explains why $|g,1,1\rangle$ is essentially occupied at all because the condition for occupation $C_{g11}\neq0$ does not identify a threshold since $C_{g11}^{(0)}\neq0$ for any nonzero drives.

Now we include the the cavity mediated cross coupling $J_{12}$ in the first order occupation which is given by 
\begin{flalign}
&C_{g20}^{(1)} = -\frac{\sqrt2\,\eta_1\eta_2 J_{12}e^{-i\phi}}{\Delta_1^2\Delta_2}, \hspace{0.5em} 
C_{g02}^{(1)} = -\frac{\sqrt2\,\eta_1\eta_2 J_{12}e^{-i\phi}}{\Delta_1\Delta_2^2}, \tag{A14}&\\
&C_{g11}^{(1)} = -\frac{J_{12}}{\Delta_1\Delta_2}\left(\frac{\eta_1^2}{\Delta_1} + \frac{\eta_2^2 e^{-2i\phi}}{\Delta_2}\right). \tag{A15}&
\end{flalign}
Note that the factors $J_1$ and $J_2$ do not appear at this order and the occupations $C_{e10},C_{e01}$ vanish at zeroth order. Consequently, the qubit-mediated pathway $|g,1,0\rangle\to|e,0,0\rangle\to|g,0,1\rangle\to|g,1,1\rangle$, which requires two qubit couplings and thus it contributes at second order only.

We carry the recursion up to second occupation which give qubit mediated contributions including the effective coupling $J_1$ and $J_2$ is given by

\begin{flalign}
\begin{split}
C_{g11}^{(2)} ={}& \frac{\eta_1\eta_2 e^{-i\phi}\big(\Delta_1 J_2^2 + \Delta_2 J_1^2 + 3\Delta_g J_{12}^2\big)}{\Delta_1^2\Delta_2^2\Delta_g} \\
&+ \frac{J_1J_2\big(\Delta_1\eta_2^2 e^{-2i\phi} + \Delta_2\eta_1^2\big)}{\Delta_1^2\Delta_2^2\Delta_g}
\end{split}\tag{A16}
\end{flalign}

\begin{flalign}
\begin{split}
C_{g20}^{(2)} ={}& \frac{\sqrt2}{\Delta_1^3\Delta_2^2\Delta_g}
\Big[\Delta_1\Delta_2\eta_1\eta_2 J_1J_2e^{-i\phi} \\
&+ \tfrac12\Delta_1\Delta_g\eta_2^2 J_{12}^2 e^{-2i\phi} + \Delta_2^2\eta_1^2 J_1^2 + \Delta_2\Delta_g\eta_1^2J_{12}^2\Big]
\end{split}\tag{A17}
\end{flalign}

\begin{flalign}
\begin{split}
C_{g02}^{(2)} ={}& \frac{\sqrt2}{\Delta_1^2\Delta_2^3\Delta_g}
\Big[\Delta_1^2\eta_2^2 J_2^2e^{-2i\phi} + \Delta_1\Delta_2\eta_1\eta_2 J_1J_2e^{-i\phi} \\
&+ \Delta_1\Delta_g\eta_2^2 J_{12}^2e^{-2i\phi} + \tfrac12\Delta_2\Delta_g\eta_1^2J_{12}^2\Big]
\end{split}\tag{A18}
\end{flalign}
Equation (A16) is the key result and it depends on $J_1$ and $J_2$ individually and hence on the ratio $d=g_{2c}/g_{1c}$ used throughout the
numerics and restoring it brings the analytical
heatmaps for $C_{g11}$, $|C_{g20}|^2$, $|C_{g02}|^2$ into agreement with
the full $8\times8$ diagonalization.


\begin{figure*}[htbp]
\centering
\includegraphics[width=\textwidth]{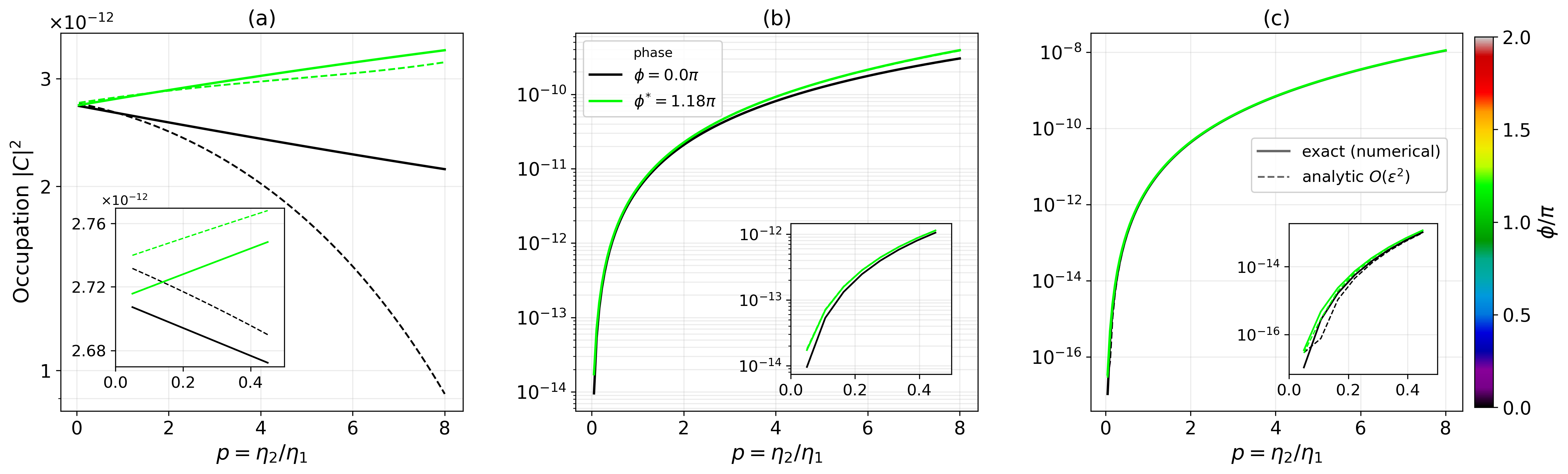}
\caption{Two-magnon occupation $|C|^2$ versus the drive ratio
$p=\eta_2/\eta_1$ for two relative phases, $\phi=0$ (black) and the optimal
phase $\phi^{*}=1.18\pi$ (green). Occupation are given as \textbf{(a)} $|C_{g20}|^2$, \textbf{(b)}
$|C_{g11}|^2$ and \textbf{(c)} $|C_{g02}|^2$ respectively.  Solid curves are the exact
solution of Eqs.~(A5)--(A9) while the dashed curves are the perturbative $O(\varepsilon^2)$
series of Eqs.~(A13)--(A18). Also the insets resolve the low-drive region near
$p\le0.5$. }
\label{fig:appendix_agreement}
\end{figure*}


Figure~\ref{fig:appendix_agreement} proofs the perturbative amplitudes of
Eqs.~(A13)--(A18) against a direct numerical solution of the $8\times8$
system at the symmetric coupling $d=1$ and a cavity detuning
$\Delta_c\simeq26.5\,g_q$ that lies comfortably inside the dispersive regime
while remaining physically representative.  The dominant amplitudes
$|C_{g11}|^2$ and $|C_{g02}|^2$ agree with the exact solution to about one per
cent at both phases.  The mode-1 amplitude $|C_{g20}|^2$ is smaller by two to
three orders of magnitude so its residual $O(\varepsilon^3)$ error is largest
in relative terms. This deviation is confined almost entirely to $\phi=0$
and at the optimal phase $\phi^{*}=1.18\pi$ all three amplitudes agree to
within about two per cent.  The agreement improves as $\Delta_c$ is increased
further.  At the operating detuning of the main text at $\Delta_c=5g_q$, the
expansion parameter is no longer small and the figures are computed from the
exact steady-state solution. Therefore, the perturbative amplitudes are retained as the
analytic expressions underlying the blockade conditions [Eq.~(A19)] and the
optimal-phase analysis.

\subsection{Combined Amplitudes and Occupation Conditions}

The full two-magnon amplitudes upto second order are
$C_{g11}\approx C_{g11}^{(0)}+C_{g11}^{(1)}+C_{g11}^{(2)}$
 from Eqs.~(A13),(A15) and (A16) and similarly for $C_{g20}$ and $C_{g02}$
from Eqs.~(A13),(A14), and (A17)--(A18) respectively . The occupation or the blockade conditions of the 
main text follow from setting each of these combined expressions to zero, i.e., $C_{g11}=0$, $C_{g20}=0$, and $C_{g02}=0$,
which are now transcendental conditions in $d=g_{2c}/g_{1c}$, $p=\eta_2/\eta_1$,
$\phi$, and the detuning or the decay ratios.

As a useful limit,we take $J_1=J_2=0$ in Eqs.~(A13)--(A15), which gives the
closed forms
\begin{flalign}
&C_{g20}=\frac{A^2}{\sqrt2\,(\Delta_1\Delta_2-J_{12}^2)^2}, 
\hspace{0.5em}  C_{g02}=\frac{B^2}{\sqrt2\,(\Delta_1\Delta_2-J_{12}^2)^2}, \notag\\
&C_{g11}=\frac{AB}{(\Delta_1\Delta_2-J_{12}^2)^2}.
\tag{A19}&
\end{flalign}
where $A\equiv\Delta_2\eta_1-J_{12}\eta_2e^{-i\phi}$ and
$B\equiv\Delta_1\eta_2e^{-i\phi}-J_{12}\eta_1$ respectively. These expressions are exact 
to all orders in $J_{12}$ in this limit and their blockade conditions $A=0$ 
and $B=0$ force $\phi=0$ or $\pi$ and coincide. This allows both same-mode 
states to be blocked precisely at one shared $p$ when 
$J_{12}^2=\Delta_1\Delta_2$. This limit is retained here only as a consistency 
check and interpretive aid which is not the full result once $J_1,J_2\neq0$.

\subsection{Correlation Functions and the CSI Violation Factor}

The equal-time second-order correlation functions are evaluated on the
truncated state (A1). The
mode occupations up to the leading order in the weak-driving expansion are $\langle m_i^\dagger m_i\rangle\simeq|C_{g10}|^2$ and
$|C_{g01}|^2$ and the two-excitation expectation values reduce to the
two-magnon amplitudes which are given by $\langle m_1^{\dagger2} m_1^{2}\rangle = 2|C_{g20}|^{2}$ , $\langle m_2^{\dagger2} m_2^{2}\rangle = 2|C_{g02}|^{2}$ and $\langle m_1^\dagger m_2^\dagger m_2 m_1\rangle = |C_{g11}|^{2}$ respectively.

The factors of $2$ in Eqs.~(A20)--(A21) arise from the doubly-excited Fock
states. Therefore, the individual correlation functions are given as
\begin{flalign}
&g_{11}^{(2)}(0) = \frac{\langle m_1^{\dagger2} m_1^{2}\rangle}{\langle m_1^\dagger m_1\rangle^{2}}
\simeq \frac{2|C_{g20}|^{2}}{|C_{g10}|^{4}}, \tag{A20}&
\end{flalign}
\begin{flalign}
&g_{22}^{(2)}(0) = \frac{\langle m_2^{\dagger2} m_2^{2}\rangle}{\langle m_2^\dagger m_2\rangle^{2}}
\simeq \frac{2|C_{g02}|^{2}}{|C_{g01}|^{4}}, \tag{A21}&
\end{flalign}
\begin{flalign}
&g_{12}^{(2)}(0) = \frac{\langle m_1^\dagger m_2^\dagger m_2 m_1\rangle}
{\langle m_1^\dagger m_1\rangle\langle m_2^\dagger m_2\rangle}
\simeq \frac{|C_{g11}|^{2}}{|C_{g10}|^{2}|C_{g01}|^{2}}. \tag{A22}&
\end{flalign}
Moreover, the individual antibunching of either Kittel mode corresponds to
$g_{ii}^{(2)}(0)<1$, i.e., suppression of the same-mode two-magnon amplitude
relative to the square of the single-magnon occupation. 

The classical Cauchy-Schwarz inequality bounds the cross-correlation by the
autocorrelations as $[g_{12}^{(2)}(0)]^{2}\le g11^{(2)}(0)\,g_{22}^{(2)}(0)$.
Therefore, its violation is quantified by the factor
\begin{flalign}
&R = \frac{[g_{12}^{(2)}(0)]^{2}}{g_{11}^{(2)}(0)\,g_{22}^{(2)}(0)}
= \frac{|C_{g11}|^{4}}{4\,|C_{g20}|^{2}|C_{g02}|^{2}},
\tag{A23}&
\end{flalign}
Here, $R>1$ is a nonclassical correlation
between the two Kittel modes that cannot be reproduced by any classical field.
Note that $R=1$ requires $|C_{g11}|^2=2|C_{g20}||C_{g02}|$ but not
$\sqrt2\,|C_{g20}||C_{g02}|$.

There are two limiting cases which provide useful observation. When the qubit is decoupled $J_1=J_2=0$ and the one and two magnon amplitudes factorize, yeilding $C_{g11}\to C_{g10}C_{g01}$,
$C_{g20}\to C_{g10}^2/\sqrt2$, and $C_{g02}\to C_{g01}^2/\sqrt2$. Hence, after
substituting these into Eq.~(A26) gives
\begin{equation}
R \big|_{J_1=J_2=0} = 1
\tag{A24}
\end{equation}
This is required by a general theorem. First, two
linearly coupled coherently driven damped oscillators generate a Gaussian
state and secondly, the gaussian states never violate the classical Cauchy-Schwarz
inequality. It follows that $R>1$ is impossible without the qubit
($J_1,J_2\neq0$) which is consistent with Eqs.~(A16)--(A18) where the qubit-dressing
terms are precise that can drive $R$ away from unity. The optimal relative phase that maximizes $R$ is obtained from the
stationarity condition $\partial_{\phi}\ln R=0$ because each two-magnon
amplitude is quadratic in $e^{-i\phi}$. This condition retains both a first
and a second harmonic in $\phi$ . The resulting optimal phase
$\phi^{*}$ is discussed in the main text.

\subsection{Optimal Relative Phase}
\label{app:optimal_phase}

The relative phase $\phi$ enters Eqs.~(A2)--(A9) only through the combination $\eta_2 e^{-i\phi}$ and it appears at most twice along any pathway to a two-magnon state in the perturbative expansion. We assume $z=e^{-i\phi}$ so that each two-magnon amplitude is an exact quadratic equation given by 
\begin{flalign}
&C_{x} = a_{x} + b_{x}\,z + c_{x}\,z^{2},
\qquad x\in\{g_{20},\,g_{11},\,g_{02}\},&
\label{eq:amp_quadratic}
\tag{A25}
\end{flalign}
where all the coefficients such as $a_x,b_x$, and $c_x$ are fixed by the couplings, detunings, and dissipation rates but independent of $\phi$. Thus, on the unit circle each $|C_x|^2$ is a trigonometric polynomial containing both first and second harmonics in $\phi$. Since $R=|C_{g11}|^{4}/\big(4|C_{g20}|^{2}|C_{g02}|^{2}\big)$ and therefore the optimal phase extremizes $\ln R$ which is given by
\begin{flalign}
&2\,\partial_{\phi}\ln|C_{g11}|^{2}
= \partial_{\phi}\ln|C_{g20}|^{2}
+ \partial_{\phi}\ln|C_{g02}|^{2}.
\label{eq:phi_star_condition}
\tag{A26}&
\end{flalign}
Equation~(\ref{eq:phi_star_condition}) is a single real equation for $\phi$ whose root maximizing $R$ gives the optimal phase $\phi^{*}$ discussed in the main text. Both harmonics contribute comparably at the operating point. Hence, Eq.~(\ref{eq:phi_star_condition}) is transcendental and has no elementary closed-form root at all. Nevertheless, solving it numerically reproduces the exact steady-state optimum to within $10^{-5}\pi$ rad over the full range of the drive ratio $p=\eta_2/\eta_1$ which confirms that the full numerical optimization is accurately captured by the above perturbative expression.

\section{Fidelity Validation of the Adiabatic Cavity Elimination}
\label{app:fidelity}

\begin{figure}[htbp]
\centering
\includegraphics[width=0.48\textwidth]{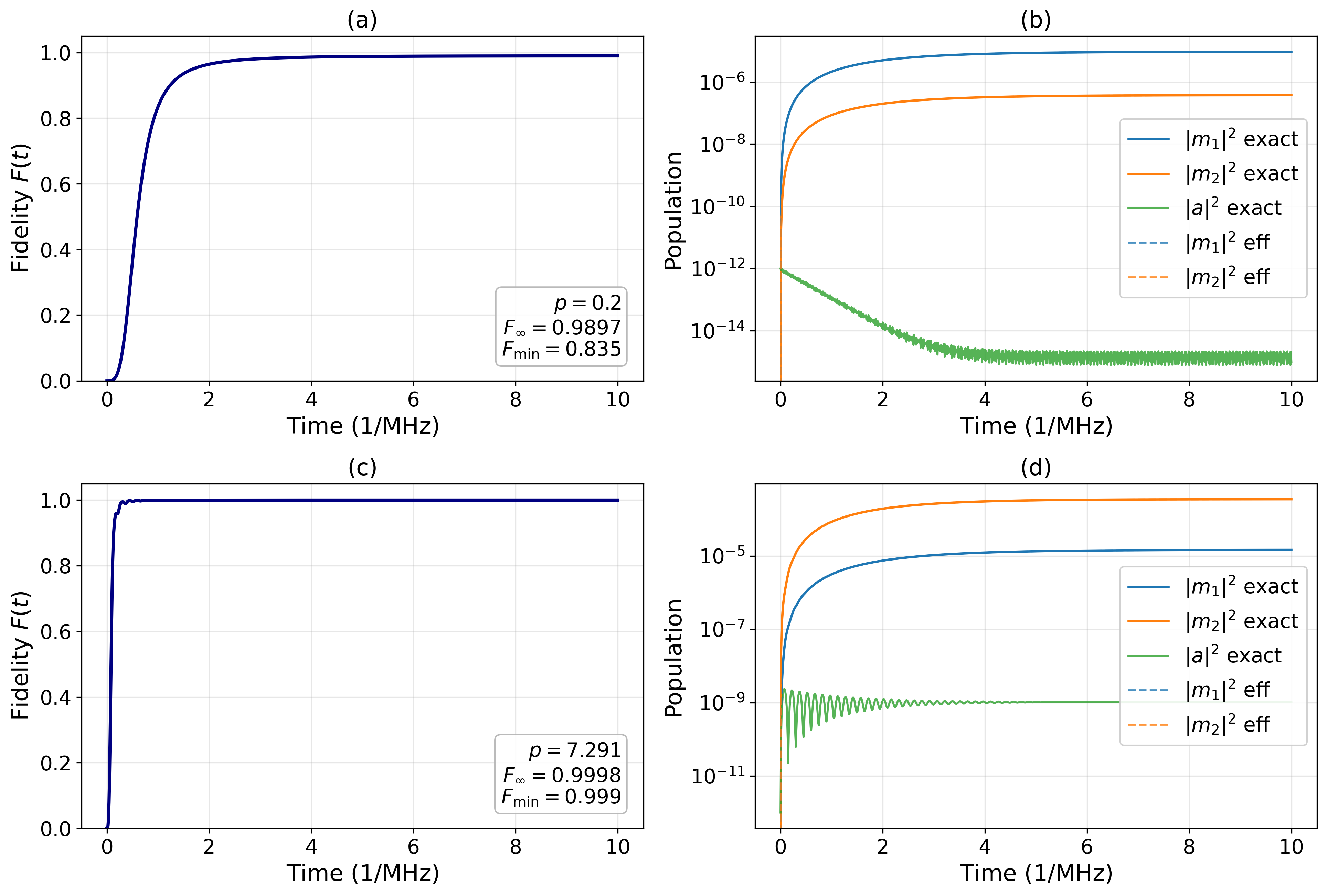}
\caption{Fidelity $F(t)$ on the left and populations on the right in a logarithmic scale of
the exact cavity-included dynamics against the effective cavity-eliminated
dynamics at the operating detuning $\Delta_c=5g_q$. \textbf{(a),(b)} at weak
drive $p=0.2$ and \textbf{(c),(d)} strong drive $p=7.291$.  Solid curves are the
exact solution and the dashed curves the effective solution respectively.}
\label{fig:Fig_Fidelity}
\end{figure}

To verify the adiabatic elimination of the cavity used to obtain the effective
Hamiltonian of Eq.~(3) we compute the fidelity between the exact dynamics by retaining the cavity  with the effective dynamics both by evolving from the ground state. The exact non-Hermitian Hamiltonian
including the cavity ($\hbar=1$) with decay rates
$\kappa_c,\kappa_1,\kappa_2 and\gamma$, is
\[
\begin{aligned}
H_{\text{exact}} =&\ \left(\Delta_c - i\tfrac{\kappa_c}{2}\right) a^\dagger a
 + \left(\Delta_{m1} - i\tfrac{\kappa_1}{2}\right) m_1^\dagger m_1 \\
&+\left(\Delta_{m2} - i\tfrac{\kappa_2}{2}\right) m_2^\dagger m_2
 + \left(\Delta_q - i\tfrac{\gamma}{2}\right) \sigma^\dagger \sigma \\
&+g_{1c}(m_1^\dagger a + m_1 a^\dagger) + g_{2c}(m_2^\dagger a + m_2 a^\dagger) \\
&+g_{qc}(\sigma^\dagger a + \sigma a^\dagger) + \eta_1(m_1^\dagger + m_1) \\
&+ \eta_2(m_2^\dagger e^{-i\phi} + m_2 e^{i\phi}),
\end{aligned}
\tag{B1}
\]
where $a$, $m_1$, $m_2$ and $\sigma$ annihilate the cavity, the two Kittel
modes and the qubit respectively.  For a large cavity--drive detuning i.e, 
$|\Delta_c|\gg g_{1c},g_{2c},g_{qc},\kappa_c$ the cavity mediates interactions while
remaining essentially unpopulated. By setting $\dot a\simeq0$ gives
\[
a \approx -\frac{g_{1c} m_1 + g_{2c} m_2 + g_{qc} \sigma}{\Delta_c - i\kappa_c/2}.
\tag{B2}
\]
So after substituting Eq.~(B2) into $H_{\text{exact}}$ reproduces the matter-only
effective Hamiltonian of Eq.~(3) with the dispersively shifted detunings
$\bar{\Delta}_i = \Delta_i - g_{ic}^2/\Delta_c - i\kappa_i/2$ and the
cavity-induced couplings $J_1=g_{1c}g_{qc}/\Delta_c$, $J_2=g_{2c}g_{qc}/\Delta_c$ and
$J_{12}=g_{1c}g_{2c}/\Delta_c$ respectively. Here, the cross-coupling $J_{12}$ is absent from
$H_{\text{exact}}$ which emerges only through the elimination.

We evolve the exact
state $|\Psi_{\text{exact}}(t)\rangle$ (basis
$|m_1,m_2,\text{qubit},\text{cavity}\rangle$) and the effective state
$|\Phi_{\text{eff}}(t)\rangle$ (basis $|m_1,m_2,\text{qubit}\rangle$) from the
ground state and compute the fidelity between the effective state and the
exact state projected onto the cavity vacuum by restricting both models to the single-excitation manifold which is given by
\[
F(t) = \frac{\big|\sum_{j} A_j^*(t)B_j(t)\big|^2}
{\big(\sum_{j}|A_j(t)|^2\big)\big(\sum_{k}|B_k(t)|^2\big)},
\tag{B3}
\]
where $A_j$ and $B_j$ are the amplitudes of the exact (cavity-projected) and
effective states in the shared matter basis.

Figure~\ref{fig:Fig_Fidelity} shows the result at the operating detuning
$\Delta_c=5g_q$ where the expansion parameter is small at around $g_q/\Delta_c=0.20$.
For strong drive at $p=7.291$, figures (c) and (d) shows that the effective description is
almost exact. The fidelity rises monotonically to $F_{\text{ss}}=0.9998$ with
no appreciable transient, and the cavity population stays suppressed to
$\sim10^{-9}$ which is several orders of magnitude below the magnon populations.  For
weak drive at $p=0.2$ figures (a) and (b) shows the steady-state fidelity is slightly
lower which is at around $F_{\text{ss}}=0.9897$ with a transient dip to $F\simeq0.84$ during
the initial turn-on.  This dip is a small-signal artifact rather than a
failure of the elimination. the magnon populations here are only
$\sim10^{-6}$ and hence the normalized fidelity is sensitive to the brief turn-on
transient while the cavity population is even more strongly suppressed at around 
$\sim10^{-15}$.  In both regimes the cavity remains essentially unpopulated
and the effective dynamics track the exact evolution in the steady state which is our exact
confirmation that the adiabatic elimination is reliable at the operating point.

\bibliography{References}

\end{document}